\documentclass[aps,prd,amsmath,floats,floatfix, twocolumn,
superscriptaddress,nofootinbib,showpacs]{revtex4-1}

\pdfoutput=1

\usepackage[colorlinks, citecolor=blue, pdfborder={0 0 0}, plainpages=false]{hyperref}
\usepackage{graphicx}
\usepackage{xspace}
\usepackage[usenames,dvipsnames]{color}
\usepackage{amssymb}
\usepackage{mathrsfs}
\usepackage{enumitem}

\allowdisplaybreaks

\usepackage{ulem}
\graphicspath{{Plots/Figures/}}

\definecolor{deepdandelion}{RGB}{215,160,65}

\newcommand{\Caltech}{\affiliation{
    TAPIR,
    Walter Burke Institute for Theoretical Physics,
    California Institute of Technology, Pasadena, California 91125, USA}}
\newcommand{\JPL}{\affiliation{Jet Propulsion Laboratory, California Institute of Technology, Pasadena,
    California 91106, USA}}

\begin{document}

\title{Spectral Cauchy-characteristic extraction of the gravitational wave news function}

\author{Kevin Barkett}\Caltech
\author{Jordan Moxon}\Caltech
\author{Mark A. Scheel} \Caltech
\author{B\'ela Szil\'agyi} \JPL\Caltech

\date{\today}

\begin{abstract}
We present an improved spectral algorithm for Cauchy-characteristic
extraction and characteristic evolution of gravitational waves in numerical relativity.  The
new algorithms improve spectral convergence both at the poles of the
spherical-polar grid and at future null infinity, as well as
increase the temporal resolution of the code. The key to the success of
these algorithms is a new set of high-accuracy tests, which we present
here.  We demonstrate the accuracy of the code and compare
with the existing \textsc{Pitt\-Null} implementation.

\end{abstract}


\maketitle

%%%%%%%%%%%%%%%%%%%%%%%%%%%%%%%%%%%%%%%%%%%%%%%%%%%%%%%%%%%%%%%%%%%%%%%%%%%%%%%

\newcommand{\vb}{\bar{v}}
\newcommand{\tb}{\bar{t}}

\section{Introduction}\label{sec:Intro}

The discovery of GW150914~\cite{LIGOVirgo2016a} heralded the beginning of
gravitational wave astronomy. In the subsequent years that detection has been
followed up by a number of other signals observed from binary black hole (BBH)
mergers~\cite{Abbott:2016nmj, Abbott:2017vtc, Abbott:2017gyy, Abbott:2017oio},
as well as from the merger of a binary neutron star (BNS)
system~\cite{TheLIGOScientific:2017qsa}. As the aLIGO~\cite{aLIGO2} and
Virgo~\cite{aVirgo2} detectors push to ever greater sensitivities, the number of
expected observations will continue to grow.

Extracting the signals from the noise involves matching the incoming data
against a template bank of theoretically expected waveforms generated across
possible binary configurations. The efficacy of extracting the configuration
parameters (for instance, masses and spins of the binary components)
from a given signal depends on the fidelity of the computed waveforms
comprising the template bank; this is because errors in the template bank will bias the
estimated parameters.
The only {\it ab initio} method of generating accurate theoretical
waveforms for merging BBH systems is via numerical relativity: the numerical
solution of the full Einstein equations on a computer.  Other methods
of generating theoretical BBH waveforms, such as
effective one-body solutions~\cite{Bohe:2016gbl} and
phenomenological models~\cite{Husa:2015iqa,Khan:2015jqa}, are calibrated to numerical relativity.

One limitation of numerical relativity
simulations is that they all rely on a Cauchy approach in which the
spacetime is decomposed into a foliation of spacelike slices, and the solution
marches from one slice to the next.  Such an approach can compute the solution to Einstein's equations only in a region
of spacetime with finite spatial and temporal
extents bounded around the compact objects,
whereas the gravitational radiation is defined at future null infinity
$\mathscr{I}^+$. While some work has gone into hyperboloidal
compactification methods for simulating the propagation of gravitational
waves to
$\mathscr{I}^+$~\cite{Husa:2005ns, Zenginoglu:2006rj, Zenginoglu2008b}, these
methods have never been fully implemented in the nonlinear regime. Without them, extracting the
waveform signal from the simulations with these
finite extents requires additional work.

The most common method of extracting the gravitational radiation from a
numerical relativity simulation is to compute quantities such as the
Newman-Penrose scalar $\Psi_4$~\cite{Newman1962} or the Regge-Wheeler and
Zerilli scalars~\cite{Sarbach2001} %~\cite{Rinne2008b}
at some large but finite distance from the near zone (perhaps 100-1000$M$,
where $M$ is the total mass of the system), typically on coordinate spheres of
constant surface area coordinate coordinate $r$.  Because these quantities or the methods of
computing them include finite-radius effects,
these quantities are computed on a series of shells at
different radii $r$, fit to a polynomial in $1/r$, and then extrapolated to infinity by
reading off the $1/r$ coefficient
of the
polynomial~\cite{Boyle-Mroue:2008}. As the extraction surfaces are shells of
constant coordinate radii, the choice of gauge implemented in the simulation can
contaminate the resulting waveforms. Furthermore, if the shells are too close to
the orbiting binary, the extrapolation procedure
might not remove all of the near-zone effects.

\begin{figure}
\includegraphics[width=.97\columnwidth]{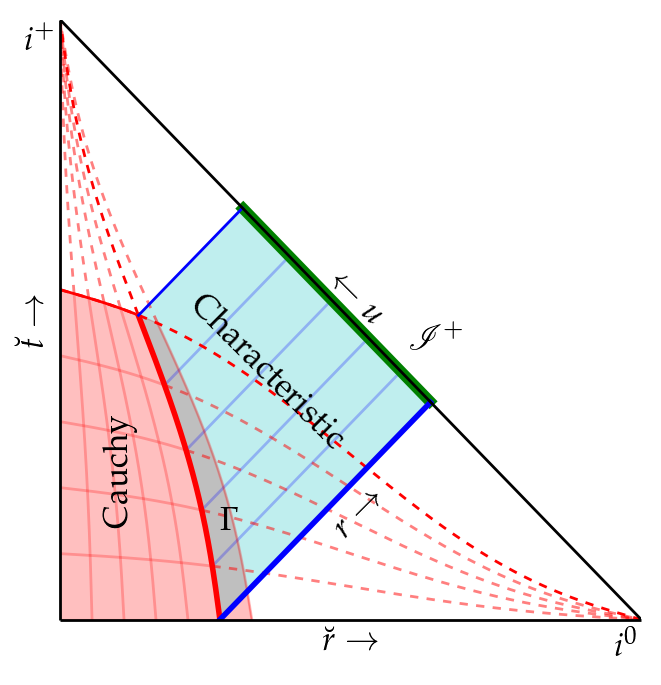}
\caption{
Penrose diagram showing a typical CCE setup. The metric is evolved using 3+1
methods in the Cauchy region (shaded red) and with null methods in the
characteristic region (shaded blue). The Cauchy and characteristic regions
overlap. Curves of constant $\breve t$ or $\breve r$, the Cauchy coordinates,
are shown in red, and are shown as dashed curves outside the Cauchy region,
where they extend to spatial infinity $i^0$ or future temporal infinity $i^+$.
Null curves of constant $u$ are shown in blue. Given data on a worldtube
$\Gamma$ (thick red curve) and on an initial null
slice (thick blue
curve), 
the characteristic evolution computes the full metric
in the characteristic region. In Sec.~\ref{sec:IBF} we describe the interface
from Cauchy to Bondi coordinates on $\Gamma$. In Sec.~\ref{sec:VE} we
describe the characteristic evolution. In Sec.~\ref{sec:SE} we discuss
computing the news function at $\mathscr{I}^+$ (thick green curve) and
transforming it to coordinates corresponding to a free-falling observer.
\label{fig:SpaceTime}
}
\end{figure}

An alternative method for computing gravitational radiation in
numerical relativity is to solve the full Einstein equations in a domain
that extends all the
way to $\mathscr{I}^+$, where gravitational waves can be measured.
This can be done by rewriting Einstein's equations using a characteristic
formalism~\cite{Bondi1962, Sachs1962, Penrose1963},
in which the equations are solved on outgoing null surfaces that extend to
$\mathscr{I}^+$.
This formalism chooses coordinates that correspond to distinct outward propagating null rays, so it fails in the dynamical, strong field regime at any location where
  outgoing null rays intersect (i.e., caustics).
Because of this, characteristic
evolution is unable to
evolve the near-field region of a merging binary system,
  so it cannot accomplish a BBH simulation on its own. However, it is possible
  to combine an interior numerical relativity code that solves the equations on
  Cauchy slices with an exterior characteristic code that solves them on null slices; the determination of characteristic quantities from Cauchy data is known as Cauchy-characteristic extraction (CCE) (see
Fig.~\ref{fig:SpaceTime}), and the subsequent numerical evolution of those quantities is known as characteristic evolution.

Specifically, CCE uses the metric and its derivatives computed from a Cauchy
  evolution (red region in Fig.~\ref{fig:SpaceTime}) and evaluated
  on a worldtube $\Gamma$ (thick red line) that
  lies on or inside the boundary of the Cauchy region.
  These quantities on the worldtube are then used as
 inner boundary data for a characteristic evolution (blue region) based on outgoing null slices (blue curves). Because the combined CCE system uses the full
Einstein equations for both the Cauchy and characteristic evolutions, it
  produces the correct solution at $\mathscr{I}^+$, with the characteristic evolution properly resolving near-zone effects.  The gravitational radiation
  is computed  according to a particular
  inertial observer at $\mathscr{I}^+$ (green curve). This observer is
related to any other inertial observer by a single Bondi-Metzner-Sachs (BMS)
transformation~\cite{Sachs1962} (the group of Lorentz boosts, rotations, and
supertranslations~\cite{Sachs1962b}),
  so up to this BMS
  transformation the waveform is independent of the gauge chosen by
the Cauchy evolution.

The first code to implement CCE and characteristic evolution was the \textsc{Pitt\-Null} code~\cite{Bishop1996,
Bishop:1997ik, Bishop1998}. Since its initial implementation there have been a
number of improvements made, and the current iteration of that code utilizes
stereographic angular coordinate patches, finite differencing, and a null
parallelogram scheme with fixed time steps for integrating in the null and time
directions. Overall the code is second-order convergent with
resolution~\cite{Reisswig:2006nt, Babiuc:2010ze}
(although a fourth-order implementation also exists, see~\cite{Reisswig:2012ka}).
Compared to waveforms computed from a Cauchy code
by evaluating $\Psi_4$ at finite radii and extrapolating to $r\to\infty$
as described above, waveforms extracted via CCE using \textsc{Pitt\-Null} were
shown to better remove gauge
effects and to better resolve
the $m=0$ memory modes~\cite{Reisswig:2009rx, Pollney:2010hs, Taylor:2013zia}.

Currently, \textsc{Pitt\-Null} requires thousands of CPU hours to compute a waveform at
$\mathscr{I}^+$ given worldtube output from a typical Cauchy BBH simulation at
multiple resolutions~\cite{Handmer:2014}. While that cost is smaller than the
computational expense of the Cauchy simulation, it is still unwieldy, and is
likely one reason that most Cauchy numerical-relativity codes do not use CCE and characteristic evolution
despite the availability of \textsc{Pitt\-Null}.
  Because the metric in the characteristic region is smooth, the computational
  cost of characteristic evolution should be greatly reduced by
  using spectral methods instead of finite differencing. Such a
spectral implementation of characteristic evolution has been introduced in the
SpEC framework~\cite{Handmer:2014, Handmer:2015, Handmer:2016}. Their tests
showed improved speed and accuracy over the finite-difference
implementation of \textsc{Pitt\-Null}~\cite{Handmer:2014, Handmer:2015}.

Our work here describes improvements in accuracy, efficiency, and robustness to
the code described in~\cite{Handmer:2014, Handmer:2015, Handmer:2016}. In
particular, we discuss an improved handling of the integration along the null
slices, we
clarify issues related to the particular choice of coordinates along the
null slice, and we implement better handling of the inertial coordinates at
$\mathscr{I}^+$. We demonstrate through a series of analytic tests that our
version of CCE and characteristic evolution can compute waveforms with much lower computational
cost than \textsc{Pitt\-Null}. An earlier version of our
  implementation has been used to probe the near-field region of a
binary black hole ringdown~\cite{Bhagwat:2017tkm}.

We start with a brief summary of the Bondi metric and the null formulation of
the Einstein equations in Sec.~\ref{sec:Bondi}. A detailed explanation for how
CCE and characteristic evolution works can be broken up into three distinct parts: the inner boundary
formalism, the volume characteristic evolution, and the $\mathscr{I}^+$
extraction, which we describe in
subsequent sections.
Section~\ref{sec:IBF} describes the means by which the metric 
known
on a worldtube is converted into Bondi form to serve as the inner boundary values for the
characteristic evolution
system. Section~\ref{sec:VE} discusses the process of
evolving Einstein's equations from the inner boundary to $\mathscr{I}^+$.
Section~\ref{sec:SE} explains how to take the metric 
 computed on $\mathscr{I}^+$ and extract the Bondi news function in the
frame of an inertial observer at $\mathscr{I}^+$. In Sec.~\ref{sec:CodeTests}, we
describe code tests and performance.

Throughout this paper, indices with Greek letters ($\mu,\nu,\ldots$) correspond to
spacetime coordinates, lowercase Roman letters ($i,j,\ldots$) to spatial
coordinates, and capitalized Roman letters ($A,B,\ldots$) to angular coordinates,
and we choose a system of geometrized units ($c=G=1$). For convenience, we have
included a definitions key in Appendix~\ref{subsec:Key}.

\section{Summary of characteristic formulation}\label{sec:Bondi}

In the characteristic region (see Fig.~\ref{fig:SpaceTime}), we adopt a
coordinate system $x^\mu=(u,r,x^A)$, where $u$ is
the
coordinate labeling the outgoing null cones, $r$ is
an areal radial
  coordinate, and $x^A$
  are the angular coordinates. Note that a curve
  of constant $(u,x^A)$
  is an outgoing null ray parametrized by $r$; for this reason we sometimes
  call $r$ a ``radinull'' coordinate. The metric can then be
expressed in the Bondi-Sachs form~\cite{Bondi1962, Sachs1962},
\begin{align}
ds^2 =& -\left(e^{2\beta}\left(1+rW\right) - r^2h_{AB}U^AU^B\right)du^2 \nonumber\\
      -& 2e^{2\beta}dudr - 2r^2h_{AB}U^Bdudx^A \nonumber\\
      +& r^2h_{AB}dx^Adx^B,
\label{eq:BondiMetric}
\end{align}
where $W$ corresponds to the mass aspect, $U^A$ to the shift, $\beta$ to the
lapse, and $h_{AB}$ to the spherical 2-metric. The quantity $h_{AB}$ has the
same determinant as the unit sphere metric $q_{AB}$, $|h_{AB}| = |q_{AB}|$.
Note that the metric Eq.~(\ref{eq:BondiMetric}) is not constrained to be
asymptotically flat, as required by Bondi-Sachs coordinates.
Instead, we impose the weaker constraint that all metric components of
Eq.~(\ref{eq:BondiMetric}) are asymptotically finite at $\mathcal{I}^+$.
To emphasize this subtle difference with Bondi-Sachs coordinates, we refer to
the spacetime metric as having the ``Bondi-Sachs form'' rather than being
expressed in Bondi-Sachs coordinates.
An additional intermediate quantity, $Q_A$, is defined to reduce
the evolution equations to a series of first order partial differential equations (PDEs),
\begin{align}
Q_A = r^2e^{-2\beta}h_{AB}U^B_{,r}.
\label{eq:QaDefinition}
\end{align}

Instead of expressing the metric in terms of tensorial objects, we employ a
complex dyad so that the metric components can be computed as
{spin-weighted scalars, and each of these scalars can be expanded
  in terms of spin-weighted
spherical harmonics (SWSHes) of the appropriate spin weight;
see Appendix~\ref{subsec:SWSH} for details about
SWSHes. The dyad $q^A$ has the following properties:
\begin{align}
q^Aq_A =& 0, \\
q^A\bar q_A =& 2.
\label{eq:Dyad}
\end{align}
If we define $q_{AB}$ and $q^{AB}$ such that
\begin{align}
  q_{AB} =& \frac{1}{2}(q_A\bar q_B+\bar q_Aq_B), \\
  q^{AC}q_{CB} =& \delta^A_B, 
\end{align}
then
\begin{align}
q^A =& q^{AB}q_B.
\end{align}

We express the metric coefficients and the quantity $Q_A$ in terms
  of spin-weighted scalars $J$, $K$, $U$, and $Q$, defined by 
\begin{align}
J =& \frac{1}{2}h_{AB}q^Aq^B, \label{eq:Jdef}\\
K =& \frac{1}{2}h_{AB}q^A\bar q^B, \label{eq:Kdef}\\
U =& q_AU^A, \label{eq:Udef}\\
Q =& Q_Aq^A.
\label{eq:Qdef}
\end{align}
The determinant condition on $h_{AB}$ defines a relationship between $J$ and $K$
as
\begin{align}
K=\sqrt{1+J\bar J}.
\end{align}
We introduce one more intermediate variable $H$,
the time derivative of $J$ along slices of constant $r$,
\begin{align}
  H = J_{,u|x^A,r=\text{const}}
  \label{eq:Hdefinition}
\end{align}
The quantities $J,\beta,$ and $Q$ are all dimensionless while $U, W,$ and $H$
have units of $1/R$ (identically, units of $1/u$ in the case of $H$).
 
Evaluating the components of the Einstein equation $G_{\mu\nu}=0$ provides a
system of
equations for
the quantities $\beta$, $Q$, $U$, $W$, and $H$:
\begin{align}
\beta_{,r} =& \mathcal{N}_\beta,
\label{eq:betaLin}\\
(r^2Q)_{,r} =& -r^2(\bar\eth J+\eth K)_{,r} + 2r^4\eth(r^{-2}\beta)_{,r}
    \nonumber\\
    +& \mathcal{N}_Q, \\
U_{,r} =& r^{-2}e^{2\beta}Q + \mathcal{N}_U, \\
(r^2W)_{,r} =& \frac{1}{2}e^{2\beta}\mathcal{R} - 1-e^\beta\eth\bar\eth e^\beta
    \nonumber\\
    +& \frac{1}{4}r^{-2}(r^4(\eth\bar U+\bar\eth U))_{,r} + \mathcal{N}_W, \\
2(rH)_{,r} =& ((1+rW)(rJ)_{,r})_{,r} - r^{-1}(r^2\eth U)_{,r} \nonumber\\
    +& 2r^{-1}e^\beta\eth^2e^\beta - (rW)_{,r}J + \mathcal{N}_J,
\label{eq:LinearizedAlphabetSoupEqns}
\end{align}
where
\begin{align}
\mathcal{R} =& 2K-\eth\bar\eth K+\frac{1}{2}(\bar\eth^2 J+\eth^2\bar J)
    +\frac{1}{4K}(\bar\eth\bar J\eth J-\bar\eth J\eth\bar J),
\label{eq:Rcurve}
\end{align}
and $\mathcal{N}_\beta,\mathcal{N}_W,\mathcal{N}_Q,\mathcal{N}_W,$ and
$\mathcal{N}_J$ are the terms nonlinear in $J$ and its derivatives, according
to~\cite{Bishop1996}.
Appendix~\ref{subsec:NonLinear} provides the full
expressions for these equations.

These equations correspond to different components of the Einstein equations,
namely, $R_{rr}=0$ gives the equation for $\beta_{,r}$,
$R_{rA}q^A=0$ gives the equation for $U_{,r}$,
$R_{AB}h^{AB}$=0 gives the equation for $W_{,r}$, and
$R_{AB}q^Aq^B=0$ gives the equation for $H_{,r}$. These
cover six of the ten independent components of Einstein's equations.
As~\cite{Bishop:1997ik}
discusses in more detail, of the four remaining components of the Einstein
equations, one of these is identically zero ($R^r_r=0$)
while the other three ($R^r_u=0$ and $R^r_Aq^A=0$) serve as constraint
conditions for the evolution on each of the null slices.

However, computing these constraint conditions involve
lengthy expressions that include the
$u$-derivatives of evolution quantities other than $J_{,u}$.
It is not straightforward to compute these derivatives to the same accuracy
achieved by the rest of the code. We leave to future implementations the ability
to accurately compute these constraints as a monitor of how well we obey the
full Einstein equations during the evolution.

The equations are presented in a useful hierarchical order: the right-hand side
of the $\beta$ equation involves only $J$ and its hypersurface derivatives, the
right-hand side of the $Q$ equation involves only $J$ and $\beta$ and their
hypersurface
derivatives, and so on for the other equations. Therefore, given data for all
quantities on the inner boundary as well as $J$ on an initial $u=$ const null
slice, we can integrate the series of equations in
Eqs.~(\ref{eq:betaLin})-(\ref{eq:LinearizedAlphabetSoupEqns})
on that slice
  from the inner boundary to $r=\infty$ to obtain $\beta, Q, U, W,$ and then $H$
  in sequence on that slice.  Then,
  given $H=J_{,u|r=\text{const}}$ on that slice, we can
integrate forward in time to obtain $J$ on the next null slice.

\section{Inner Boundary Formalism}\label{sec:IBF}

The coordinates used to evolve Einstein's equations in the Cauchy
  region (red area of Fig.~\ref{fig:SpaceTime}) are generally different
  from the coordinates discussed in Sec.~\ref{sec:Bondi}. The
  Cauchy coordinates are chosen to make the interior evolution proceed without
  encountering coordinate singularities; the procedure for choosing these
  coordinates is complicated and typically involves coordinates that are
  evolved along with the
  solution~\cite{Pretorius2005a, Lindblom2006, Szilagyi:2009qz, Scheel2009,
    Campanelli2006a, Baker2006b}.
 Therefore, for CCE we must transform from arbitrary Cauchy coordinates to
  coordinates such that the spacetime metric takes
    the Bondi-Sachs form [Eq.~(\ref{eq:BondiMetric})] at the worldtube.

Here, in the Cauchy region, for simplicity we assume Cartesian coordinates
 $(\breve t, \breve x^{\breve\i})$ in which the worldtube hypersurface $\Gamma$
 (which is chosen by the Cauchy code) is a surface of constant $\breve r$, where
 $\breve r = \sqrt{\breve x^2+\breve y^2+\breve z^2}$.

We also define angular coordinates
 $\breve x^{\breve A} = (\breve\theta,\breve\phi)$ in the usual way from the
 Cartesian coordinates $\breve x^{\breve\i}$.

The worldtube serves as the inner boundary of the characteristic domain (see
Fig.~\ref{fig:SpaceTime}). On this boundary, we assume that
the interior Cauchy code provides the spatial
3-metric $g_{\breve\i\breve\j}$, the shift $\beta^{\breve\i}$, and
the lapse $\breve\alpha$, along with the $\breve r$ and $\breve t$ derivatives of each of these quantities.
Angular derivatives of these quantities are necessary as well; however, we can
compute those numerically within the worldtube itself,
  so they need not be provided
{\it a priori}.

Reference~\cite{Bishop1998} describes how to take
  the data provided by the interior Cauchy code and
covert it into Bondi form [Eq.~(\ref{eq:BondiMetric})] to extract the inner
boundary values of the evolution quantities ($J_{|\Gamma}, \beta_{|\Gamma},
\ldots$). This section is primarily a summary of their results; however, we
use different notation than Ref.~\cite{Bishop1998}.
Additionally, as noted above,
the SpEC CCE treatment takes the inner boundary of the domain
to be the worldtube provided by the Cauchy code,
which is generally not a surface of constant $r$. The \textsc{Pitt\-Null}
treatment, on the other hand, uses a surface of constant $r$ as
the inner boundary of the domain, and
performs a Taylor expansion in the affine radial coordinate in order
to determine inner boundary data on this surface.
Avoiding the Taylor expansion simplifies the boundary computation and may
provide marginal precision improvements by avoiding a finite Taylor series
truncation error.

\subsection{Affine null coordinates}

Our goal is to transform from the coordinates $(\breve t, \breve x^{\breve\i})$
to coordinates such that the metric takes the
  Bondi-Sachs form (Eq.~(\ref{eq:BondiMetric})).
It is simplest to proceed in two steps: the first step, described in this
subsection, is to construct coordinates foliated by outgoing null geodesics.
The second step, described in Sec.~\ref{sec:bondi-form-metric}, will be to
transform from these affine coordinates to Bondi coordinates.

We begin by constructing a choice null generator $\ell^{\breve\mu}$,
which involves
the unit outward spatial vector normal to the worldtube's surface,
$s^{\breve\mu}$, and the unit timelike vector normal to a slice of constant
$\breve t$, $n^{\breve\mu}$:
\begin{align}
s^{\breve\mu} =& \left\{0,\frac{g^{\breve\i\breve\j}\breve x_{\breve\j}}
    {\sqrt{g^{\breve\i\breve\j}\breve x_{\breve\i}\breve x_{\breve\j}}}\right\},
    \label{eq:WTOutwardNormal}\\
n^{\breve\mu} =& \frac{1}{\breve\alpha}\left\{1,-\beta^{\breve\i}\right\}.
    \label{eq:WTTimeVector}
\end{align}
Equation~(\ref{eq:WTOutwardNormal}) depends on our simplifying assumption
that the worldtube is spherical in Cauchy coordinates, and can be generalized.
From these equations, the null generator is
\begin{align}
\ell^{\breve\mu} =& \frac{n^{\breve\mu}+s^{\breve\mu}}
    {\breve\alpha - g_{\breve\i\breve\j}\beta^{\breve\i}s^{\breve\j}}.
    \label{eq:WTNullGenerator}
\end{align}
The time derivatives of these vectors are
\begin{align}
s^{\breve\mu}_{,\breve t} =& \left\{0,
    ~(-g^{\breve\i\breve\j}+s^{\breve\i}s^{\breve\j}/2)
    s^{\breve k}g_{\breve\j\breve k,\breve t}\right\}, \\
n^{\breve\mu}_{,\breve t} =& \frac{1}{\breve\alpha^2}
    \left\{-\breve\alpha_{,\breve t},~\breve\alpha_{,\breve t}\beta^{\breve\i}
    - \breve\alpha\beta^{\breve\i}_{,\breve t}\right\}, \\
\ell^{\breve\mu}_{,\breve t} =& \frac{n^{\breve\mu}_{,\breve t}
    + s^{\breve\mu}_{,\breve t} + \ell^{\breve\mu}\left(-\breve\alpha_{,\breve t}
    + g_{\breve\i\breve\j,\breve t}\beta^{\breve\i}s^{\breve\j}
    + g_{\breve\i\breve\j}\beta^{\breve\i}_{,\breve t}s^{\breve\j}
    + g_{\breve\i\breve\j}\beta^{\breve\i}s^{\breve\j}_{,\breve t}\right)}
    {\breve\alpha - g_{\breve\i\breve\j}\beta^{\breve\i}s^{\breve\j}}.
\end{align}

We will now construct a null coordinate system based on outgoing null geodesics
generated by $\ell^{\breve\mu}$. Let $\bar\lambda$ be an affine parameter along
these geodesics such that the value of $\bar\lambda$ on the worldtube $\Gamma$ is
$\bar\lambda_{|\Gamma}=0$. We also define a null coordinate $\bar u$ and angular
coordinates $\bar x^{\bar A}=(\bar\theta,\bar\phi)$ that obey $\bar u =\breve t$
and $\bar x^{\bar A} = \breve x^{\breve A}$ on the worldtube, and are constant
along the outgoing null geodesic generated by $\ell^{\breve\mu}$. Thus we have
defined a new intermediate, affine coordinate system, $\bar x^{\bar\mu} = (\bar u,
\bar\lambda, \bar\theta, \bar\phi)$, and we will express the metric
$g_{\bar\mu\bar\nu}$ in these affine coordinates.

To do this, we will need to write down the coordinate transformation
  from $\breve x^{\breve\mu}$ to $\bar x^{\bar\mu}$ in a
  neighborhood of the worldtube, not just on the worldtube itself, because
  we need derivatives of this transformation.  In particular, we will need
  derivatives with respect to $\bar\lambda$.
  The derivative of the
  metric components $g_{\breve\mu\breve\nu}$ along
the null direction simply is
\begin{align}
g_{\breve\mu\breve\nu,\bar\lambda} =& \ell^{\breve\gamma}g_{\breve\mu\breve\nu,\breve\gamma}.
\end{align}
The evolution of the coordinates $\breve x^{\breve\mu}$ along null geodesics
implies that in a neighborhood of the worldtube
\begin{align}
\breve x^{\breve\mu}_{,\bar\lambda} =
    \ell^{\breve\nu}\partial_{\breve\nu}\breve x^{\breve\mu}
    = \ell^{\breve\mu}.
\label{eq:DbrevexDlambda}
\end{align}

Given the new coordinates $\bar x^{\bar\mu}$, the metric components in these
coordinates are
\begin{align}
g_{\bar\mu\bar\nu} = \frac{\partial\breve x^{\breve\alpha}}{\partial \bar x^{\bar\mu}}
    \frac{\partial\breve x^{\breve\beta}}{\partial \bar x^{\bar\nu}}g_{\breve\alpha\breve\beta}.
    \label{eq:NullMetricDef}
\end{align}

On the worldtube,
\begin{align}
  \frac{\partial \breve{t}}{\partial \bar{x}^{\bar{A}}} &= 0 \nonumber\\
  \frac{\partial \breve{x}^{\breve{i}}}{\partial \bar{x}^{\bar{A}}} &= \frac{\partial \breve{x}^{\breve{i}}}{\partial \breve{x}^{\breve{A}}}\nonumber\\
\frac{\partial\breve t}{\partial\bar u} &= 1, \nonumber\\
\frac{\partial\breve x^{\breve\i}}{\partial\bar u} &= 0,
\end{align}
where the term $\partial\breve x^{\breve\i}/\partial\bar x^{\bar A}$
is the standard Cartesian to spherical Jacobian. The above values of the Jacobians
hold only on the worldtube. In addition to the metric itself, we will also need
first derivatives of the metric, including the derivative with respect to
$\bar\lambda$. This requires the $\bar\lambda$ derivatives of the Jacobians evaluated on
the worldtube, which we represent here as
\begin{align}
\frac{\partial^2\breve x^{\breve\mu}}{\partial\bar x^{\bar A}\partial\bar\lambda} =&
  \frac{\partial\ell^{\breve\mu}}{\partial\bar x^{\bar A}} = \ell^{\breve\mu}_{,\bar A}, \nonumber\\
\frac{\partial^2\breve x^{\breve\mu}}{\partial\bar u\partial\bar\lambda} =&
  \frac{\partial\ell^{\breve\mu}}{\partial\bar u} = \ell^{\breve\mu}_{,\bar u},
\end{align}
where we have made use of Eq.~(\ref{eq:DbrevexDlambda}).

We are now ready to write out
the metric in these intermediate coordinates by taking the expression
in Eq.~(\ref{eq:NullMetricDef}) and taking the appropriate derivatives,
\begin{align}
g_{\bar u\bar\lambda} =& -1, \nonumber\\
g_{\bar\lambda\bar\lambda} =& g_{\bar\lambda\bar A} = 0, \nonumber\\
g_{\bar u\bar u} =& g_{\breve t\breve t}, \nonumber\\
g_{\bar u\bar A} =& \frac{\partial\breve x^{\breve\i}}{\partial\bar x^{\bar A}}
    g_{\breve\i\breve t}, \nonumber\\
g_{\bar A\bar B} =& \frac{\partial\breve x^{\breve\i}}{\partial\bar x^{\bar A}}
    \frac{\partial\breve x^{\breve\j}}{\partial\bar x^{\bar B}}g_{\breve\i\breve\j}, \nonumber\\
g_{\bar A\bar B,\bar\lambda} =& \frac{\partial\breve x^{\breve\i}}{\partial\bar x^{\bar A}}
    \frac{\partial\breve x^{\breve\j}}{\partial\bar x^{\bar B}}g_{\breve\i\breve\j,\bar\lambda} \nonumber\\
    +& \left(\ell^{\breve\mu}_{~,\bar A}\frac{\partial\breve x^{\breve\i}}{\partial\bar x^{\bar B}}
    + \ell^{\breve\mu}_{~,\bar B}\frac{\partial\breve x^{\breve\i}}{\partial\bar x^{\bar A}}\right)
    g_{\breve\mu\breve\i}, \nonumber\\
g_{\bar A\bar B,\bar u} =& \frac{\partial\breve x^{\breve\i}}{\partial\bar x^{\bar A}}
    \frac{\partial\breve x^{\breve\j}}{\partial\bar x^{\bar B}}g_{\breve\i\breve\j,\breve t}, \nonumber\\
g_{\bar u\bar A,\bar\lambda} =& \ell^{\breve\mu}_{~,\bar A}g_{\breve t\breve\mu}
    + \frac{\partial\breve x^{\breve\i}}{\partial\bar x^{\bar A}}\left(g_{\breve\i\breve t,\bar\lambda}
    + \ell^{\breve\mu}_{~,\bar u}g_{\breve\i\breve\mu}\right),
\end{align}
and
\begin{align}
g^{\bar u\bar u} =& g^{\bar u\bar A} = 0, \nonumber\\
g^{\bar u\bar\lambda} =& -1, \nonumber\\
g^{\bar A\bar B}g_{\bar B\bar C} =& \delta^{\bar A}_{\bar C}, \nonumber\\
g^{\bar\lambda\bar A} =& g^{\bar A\bar B}g_{\bar u\bar B}, \nonumber\\
g^{\bar\lambda\bar\lambda} =& -g_{\bar u\bar u} + g^{\bar\lambda\bar A}g_{\bar u\bar A}, \nonumber\\
g^{\bar A\bar B}_{~~~,\bar\lambda} =& -g^{\bar A\bar C}g^{\bar B\bar D}g_{\bar C\bar D,\bar\lambda}, \nonumber\\
g^{\bar\lambda\bar A}_{~~~,\bar\lambda} =& g^{\bar A\bar B}\left(g_{\bar u\bar B,\bar\lambda}
    - g^{\bar\lambda\bar C}g_{\bar B\bar C,\bar\lambda}\right).
\label{eq:InterCoordsMet}
\end{align}

\subsection{Bondi form of metric}
\label{sec:bondi-form-metric}

Given the intermediate null coordinates and the metric in that coordinate
system, we apply one last coordinate transformation to 
put the spacetime metric in Bondi-Sachs form (Eq.~(\ref{eq:BondiMetric})).
We define coordinates $(u,r,\theta,\phi)$, where $r$ is a surface area
coordinate, $u=\bar u$, $\theta=\bar\theta$, and $\phi = \bar\phi$. The surface
area coordinate $r$ is defined by
\begin{equation}
    r = \left(\frac{|g_{AB}|}{|q_{AB}|}\right)^{\frac{1}{4}}
    = \left(\frac{|g_{\bar A\bar B}|}{|q_{\bar A\bar B}|}\right)^{\frac{1}{4}},
    \label{eq:BondiRadius}
\end{equation}
where $q_{\bar A\bar B}$ is the unit sphere metric.

The components of the metric in Bondi coordinates are then
\begin{align}
g^{\mu\nu} = \frac{\partial x^\mu}{\partial\bar x^{\bar\alpha}}
\frac{\partial x^\nu}{\partial\bar x^{\bar\beta}}g^{\bar\alpha\bar\beta}.
\label{eq:BondiMetricTransformation}
\end{align}
The Jacobians include the derivatives of the surface area coordinate $r$.
We compute
\begin{align}
           r_{,\bar\alpha} =& \frac{r}{4}\left(g^{\bar A\bar B}g_{\bar A\bar B,\bar\alpha}
                              - \frac{|q_{\bar{A}\bar{B}}|_{, \bar{\alpha}}}{|q_{\bar{A} \bar{B}}|}\right).
\label{eq:BondiRDerivs}
\end{align}

Since the only difference between the final boundary coordinates
  $(u, r, \theta, \phi)$ and intermediate coordinates is the
choice of radinull coordinates, the Jacobians for the $u$, $\theta$, and $\phi$
directions are trivial. Equation~(\ref{eq:InterCoordsMet}) gives us
\begin{align}
g^{uu} =& g^{\bar u\bar u} = 0, \nonumber\\
g^{uA} =& g^{\bar u\bar A} = 0, \nonumber\\
g^{AB} =& g^{\bar A\bar B}.
\end{align}

The other metric components are
\begin{align}
g^{ur} =& \frac{\partial r}{\partial\bar x^{\bar\mu}}g^{\bar u\bar\mu}
    = -r_{,\bar\lambda}, \nonumber\\
g^{rr} =& \frac{\partial r}{\partial\bar x^{\bar\mu}}
    \frac{\partial r}{\partial\bar x^{\bar\nu}}g^{\bar\mu\bar\nu}
    = \left(r_{,\bar\lambda}\right)^2
    g^{\bar\lambda\bar\lambda} \nonumber\\
    +& 2r_{,\bar\lambda}\left(
    r_{,\bar A}g^{\bar\lambda\bar A}
    - r_{,\bar u}\right)
    + r_{,\bar A}r_{,\bar B}g^{\bar A\bar B}, \nonumber\\
g^{rA} =& \frac{\partial r}{\partial\bar x^{\bar\mu}}g^{\bar A\bar\mu}
    = r_{,\bar\lambda}g^{\bar\lambda\bar A} + r_{,\bar B}g^{\bar A\bar B}.
\end{align}

From this we can also construct the inverse Jacobian elements. The
elements of that Jacobian we shall need are
\begin{align}
\frac{\partial\bar u}{\partial u} &= 1, \nonumber\\
\frac{\partial\bar u}{\partial x^i} &= 0, \nonumber\\
\frac{\partial\bar\lambda}{\partial u} &= -\frac{r_{,\bar u}}{r_{,\bar\lambda}}, \nonumber\\
\frac{\partial\bar x^{\bar A}}{\partial x^A} &= \delta^{\bar A}_{A}. \nonumber\\
\frac{\partial\bar x^{\bar A}}{\partial r} = \frac{\partial\bar x^{\bar A}}{\partial u} &= 0.
\end{align}
The final metric element we shall want is $g_{AB}$ which we can compute as
\begin{align}
g_{AB} &= \frac{\partial\bar x^{\bar\alpha}}{\partial x^A}
    \frac{\partial\bar x^{\bar\beta}}{\partial x^B} g_{\bar\alpha\bar\beta} \nonumber\\
    &= g_{\bar A\bar B} + \frac{\partial\bar\lambda}{\partial x^B} g_{\bar\lambda\bar A}
    + \frac{\partial\bar\lambda}{\partial x^A} g_{\bar\lambda\bar B}
    + \frac{\partial\bar\lambda}{\partial x^A}\frac{\partial\bar\lambda}{\partial x^B}
    g_{\bar\lambda\bar\lambda} \nonumber\\
    &= g_{\bar A\bar B}
\end{align}
where we made use of the fact that $g_{\bar\lambda\bar\lambda} =
g_{\bar\lambda\bar A} = 0$.

Because $u$ and $x^A$
  are equal to $\breve t$ and $\breve x^{\breve A}$ on the worldtube
  and are constant along outgoing null geodesics,
  the time and angular coordinates $(\breve t,\breve x^{\breve A})$
  on the worldtube
  determine the coordinates $u$ and $x^{A}$ throughout
  the characteristic region, including on $\mathscr{I}^+$.  Thus, the
  coordinates at $\mathscr{I}^+$ will be gauge-dependent, since
  $\breve t$ and $\bar x^{\bar A}$ are dependent upon the gauge choices made in the 3+1 Cauchy evolution.  We will
  later eliminate this gauge dependence by evolving and transforming to
  the coordinates of free-falling observers
  on $\mathscr{I}^+$, as described below in Sec.~\ref{subsec:InerCoords}.

\subsection{Inner boundary values of characteristic variables}

Now that we have the full metric in Bondi-Sachs form
  [Eq.~(\ref{eq:BondiMetric})], we assemble the inner boundary values for the
various evolution variables used in the volume, $J, \beta, Q, U, W,$ and $H$. We
write out the complex dyads as
\begin{align}
q_A =& \left\{-1, -i\sin\theta\right\}, \nonumber\\
q^A =& \left\{-1, -\frac{i}{\sin\theta}\right\}.
\label{eq:ComplexDyads}
\end{align}
Because of the identification between the intermediate angular coordinates
  $\bar{x}^{\bar{A}}$ and the characteristic coordinates $x^A$, the dyads are
  identified, $q^A = q^{\bar A}$ and $q_A = q_{\bar A}$. Then, as a consequence of Eq.~(\ref{eq:ComplexDyads}),
  $q_{A,\bar\lambda}=q^A_{~,\bar\lambda}=0$ and $q_{A,\bar u}=q^A_{~,\bar u}=0$.

Inverting the metric in Eq.~(\ref{eq:BondiMetric}),
\begin{align}
g^{\mu\nu} =
  \begin{bmatrix}
  0 & -e^{-2\beta} & 0^A \\
  -e^{-2\beta} & (1+rW)e^{-2\beta} & -e^{-2\beta}U^A \\
  0^B & -e^{-2\beta}U^B & r^{-2}h^{AB}
  \end{bmatrix},
\end{align}
where $h_{AB}h^{BC}=\delta^C_A$ and $|h_{AB}|=|q_{AB}|$.

In the \textsc{Pitt\-Null} code,
the quantities $J$, $\beta$, $Q$, $U$, and $W$ and their
$\bar\lambda$ derivatives are computed
using an expansion in affine coordinates to compute their values along
a surface of constant surface area coordinate $r$~\cite{Bishop1998}.
  \textsc{Pitt\-Null} then chooses its internal compactified radinull
  coordinates in the
  characteristic region to be
surfaces of constant $r$.
However, in Ref.~\cite{Handmer:2014} and here, we choose
our inner boundary to be the worldtube. The value of the surface area coordinate $r$
at the worldtube we define as $R(u,x^A)$,
\begin{align}
R =& r_{|\Gamma}, \label{eq:WorldTubeBondiRadius} \\
R_{,\bar\lambda} =& r_{,\bar\lambda|\Gamma},\\
R_{,\bar u} =& r_{,\bar u|\Gamma}\label{eq:WTBondiRDerivU}.
\end{align}
The consequences of this change in the inner boundary hypersurface are discussed in
more detail within Sec.~\ref{sec:VECD}.

We can now write down the inner boundary values of the
  characteristic variables in terms of the metric coefficients that we have
  computed at the inner boundary.
Going back to the definition of $J=\frac{1}{2}q^Aq^Bh_{AB}$, we get the
expressions
\begin{align}
J_{|\Gamma} =& \frac{1}{2R^2}q^Aq^Bg_{AB} = \frac{1}{2R^2}q^{\bar A}q^{\bar B}g_{\bar A\bar B}, \label{eq:JIB}\\
K_{|\Gamma} =& \sqrt{1 + J_{|\Gamma}\bar J_{|\Gamma}}, \\
J_{,\bar\lambda|\Gamma} =& \frac{1}{2R^2}q^{\bar A}q^{\bar B}g_{\bar A\bar B,\bar\lambda}
    - \frac{2R_{,\bar\lambda}}{R}J_{|\Gamma}, \label{eq:JIBlambda} \\
J_{,\bar u|\Gamma} =& \frac{1}{2R^2}q^{\bar A}q^{\bar B}g_{\bar A\bar B,\bar u}
    - \frac{2R_{,\bar u}}{R}J_{|\Gamma}.
\end{align}
To get the inner boundary value of $H$, we expand $J_{,u}$ as
\begin{align}
J_{,u} &= \frac{\partial\bar u}{\partial u}J_{,\bar u}
    + \frac{\partial\bar\lambda}{\partial u}J_{,\bar\lambda}
\end{align}
so then we find after substituting and simplifying that,
\begin{align}
H_{|\Gamma} &= \frac{1}{2R^2}q^{\bar A}q^{\bar B}\left(g_{\bar A\bar B,\bar u}
    - \frac{R_{,\bar u}}{R_{,\bar\lambda}}g_{\bar A\bar B,\bar\lambda}\right).
\end{align}

We can read off the value for $g^{ur}$ to compute $\beta$,
\begin{align}
\beta_{|\Gamma} = -\frac{1}{2}\ln\left(R_{,\bar\lambda}\right).
\label{eq:betaIB}
\end{align}
We will also need $\beta_{,\bar\lambda|\Gamma}$ in order to compute $Q_{|\Gamma}$.
Directly differentiating Eq.~(\ref{eq:betaIB}) yields
\begin{align}
  \beta_{,\bar\lambda|\Gamma} = -\frac{R_{,\bar\lambda\bar\lambda}}{2R_{,\bar\lambda}},
\label{eq:betalambdaIB1}
\end{align}
but this involves the quantity $R_{,\bar\lambda\bar\lambda}$, which appears to depend on
second derivatives of the metric. So we instead compute
$\beta_{,\bar\lambda|\Gamma}$ using $\beta$'s evolution equation,
Eq.~(\ref{eq:betaEvo}):
\begin{align}
  \beta_{,\bar\lambda|\Gamma}    
    = \frac{R}{8R_{,\bar\lambda}}\left(J_{,\bar\lambda|\Gamma}\bar J_{,\bar\lambda|\Gamma}
    - \left(K_{,\bar\lambda|\Gamma}\right)^2\right),
\label{eq:betalambdaIB}
\end{align}
which involves only first derivatives.

The quantities $U$ and $W$ can similarly be read off from the metric:
\begin{align}
U_{|\Gamma} =& \frac{g^{rA}}{g^{ur}}q_A, \\
W_{|\Gamma} =& \frac{1}{R}\left(-\frac{g^{rr}}{g^{ur}} - 1\right).
\label{eq:UWIB}
\end{align}

To get $Q_{|\Gamma}$, we will also need $U_{,\bar\lambda|\Gamma}$, which we compute
by differentiating the expression for $U_{|\Gamma}$ and using
Eq.~(\ref{eq:betalambdaIB1}) to eliminate $R_{,\bar\lambda\bar\lambda}$ in favor of
$\beta_{,\bar\lambda|\Gamma}$:
\begin{align}
U_{,\bar\lambda|\Gamma} =& -\left(g^{\bar\lambda\bar A}_{~~~,\bar\lambda}
    + \frac{R_{,\bar\lambda\bar B}}{R_{,\bar\lambda}}g^{\bar A\bar B}
    + \frac{R_{,\bar B}}{R_{,\bar\lambda}}g^{\bar A\bar B}_{~~~,\bar\lambda}
    \right)q_{\bar A} \nonumber\\
    +& 2\beta_{,\bar\lambda|\Gamma}\left(U_{|\Gamma} + g^{\bar\lambda\bar A}q_{\bar A}\right),
\label{eq:UlambdaIB}
\end{align}
where it is understood that $\beta_{,\bar\lambda|\Gamma}$ is to be evaluated using
Eq.~(\ref{eq:betalambdaIB}). Now that we have an expression for
$U_{,\bar\lambda|\Gamma}$, the inner boundary value of $Q$ is given by
\begin{align}
Q_{|\Gamma} =& R^2\left(J_{|\Gamma}\bar U_{,\bar\lambda|\Gamma}
    + K_{|\Gamma}U_{,\bar\lambda|\Gamma}\right).
\label{eq:QIB}
\end{align}

\subsection{Computational domain}

We implement angular basis functions through the use of the external code
packages {\tt Spherepack}~\cite{Boyd1989,spherepack-home-page}, which can handle
standard spherical harmonics, and {\tt Spinsfast}~\cite{Huffenberger:2010},
which is capable of handling SWSHes.
The worldtube metric and most of the intermediate quantities of the inner
boundary formalism are real, tensorial metric quantities (i.e. representable by
the typical spherical harmonics), so we use {\tt Spherepack}.
Once all of the inner boundary values of the Bondi evolution quantities
are computed, they are then projected onto the basis utilized by {\tt Spinsfast} for
use during the volume evolution. Because Cauchy codes evaluate the worldtube
data at discrete time slices, we use cubic interpolation to evaluate each of the
metric quantities at arbitrary time values.

\section{Volume Evolution}\label{sec:VE}

\subsection{Computational domain}\label{sec:VECD}

Because the domain of characteristic evolution extends all of the way out to $\mathscr{I}^+$ where
the surface area coordinate $r$ is infinite, to express $\mathscr{I}^+$ on a finite
computational domain, we
define a compactified coordinate, $\rho$,
\begin{align}
\rho=\frac{r}{R+r}
\label{eq:rho}
\end{align}
where $R$ is the surface area coordinate of the worldtube given in
Eq.~(\ref{eq:WorldTubeBondiRadius}) so that $\rho$ runs from
$\rho_{|\Gamma}=1/2$ to $\rho_{|\mathscr{I}^+}=1$.
This choice of compactification is subtly different from that which is
used in \textsc{Pitt\-Null}~\cite{Reisswig:2012ka}. Because they expand in affine
coordinates to obtain
a hypersurface of constant Bondi radius, their
compactification parameter is constant and unchanging during their entire
evolution. By tying our compactification parameter to a fixed Cauchy
coordinate radius
$\breve r$ and allowing the surface area coordinate $r$ to change freely, we must be careful in
how we define our derivatives.

One consequence of utilizing $\rho$ is that angular derivatives computed numerically on
our grid, $\eth_{|\rho}$, are evaluated at a constant value of $\rho$,
  so these are not the same as
angular derivatives defined on
surfaces of constant $r$, which we denote as $\eth$.
Since Eqs.~(\ref{eq:betaLin})--(\ref{eq:LinearizedAlphabetSoupEqns})
involve $\eth$ and not $\eth_{|\rho}$, we must apply a correction
factor to compute $\eth$ from $\eth_{|\rho}$:
\begin{align}
\eth F =& \eth_{|\rho}F - F_{,\rho}\eth_{|\rho}\rho
    = \eth_{|\rho}F - F_{,\rho}\frac{\rho(1-\rho)}{R}\eth_{|\rho}R,
\label{eq:EthCorrection}
\end{align}
  for an arbitrary spin-weighted scalar quantity $F$.
  Similar correction factors are needed for second derivatives that appear
  in the evolution equations:
\begin{widetext}
\begin{align}
  \label{eq:EthRhoCorrection}
(\eth F)_{,\rho} =& \eth_{|\rho}F_{,\rho}
    - F_{,\rho}\frac{1-2\rho}{R}\eth_{|\rho}R
    - F_{,\rho\rho}\frac{\rho(1-\rho)}{R}\eth_{|\rho}R, \\
\bar\eth\eth F =& \bar\eth_{|\rho}\eth_{|\rho}F
    + F_{,\rho}\left(\frac{\rho(1-\rho)}{R^2}\right)
    \left(2(1-\rho)\bar\eth_{|\rho}R\eth_{|\rho}R
    - R\bar\eth_{|\rho}\eth_{|\rho}R\right)
    - \eth_{|\rho}F_{,\rho}\left(\frac{\rho(1-\rho)}{R}\bar\eth_{|\rho}R\right)
    \nonumber\\
    -& \bar\eth_{|\rho}F_{,\rho}\left(\frac{\rho(1-\rho)}{R}\eth_{|\rho}R\right)
    + F_{,\rho\rho}\left(\frac{\rho(1-\rho)}{R}\right)^2
    \bar\eth_{|\rho}R\eth_{|\rho}R.
  \label{eq:EthbarEthCorrection}
\end{align}
\end{widetext}
Correction factors for $\bar\eth F$, $\bar\eth F_{,\rho}$, $\eth\eth F$, $\eth\bar\eth F$, and $\bar\eth\bar\eth F$ are obtained by appropriately
interchanging $\eth$ and $\bar\eth$
  in Eqs.~(\ref{eq:EthCorrection})-(\ref{eq:EthbarEthCorrection}).

  Numerical derivatives with respect to $t$ and $u$ are also taken at
constant $\rho$ on our grid, but at constant $r$ in the equations, so similar correction factors are required there as well, as discussed below in Sec.~\ref{subsec:TimeEvolution}.

We employ computational grid meshes suitable for spectral methods,
Chebyshev-Gauss-Lobatto for the radinull direction and {\tt Spinsfast} mesh for
the angular directions with uniform $\phi$ and $\theta$ grids.

\subsection{Spectral representability} \label{sec:SpectralRepresentability}

Spectral techniques represent functions over a finite numerical domain as a
series of polynomial functions.
Such representations are of greatest use when the numerical evolution gives rise
to smooth solutions, which converge exponentially with resolution in the
spectral expansion.
However, any defect in the solution, such as discontinuities, corners, cusps, or
the presence of logarithmic dependence, will spoil the exponential convergence
of a spectral method, and potentially introduce spurious oscillatory
contributions to the numerical result.
For this reason, it is of great importance to the characteristic evolution code
in SpEC to minimize or eliminate sources of such nonregular contributions to
the hypersurface equations.

The nature of the characteristic hypersurface equations permits
 terms proportional to $\log(r)$ to develop in the solution of the characteristic evolution system. These terms are not representable by polynomial expansions in $1/r$ or by polynomial expansions in $\rho$, so if present they spoil exponential convergence.  Such terms creep
  into the evolved solutions by three principal avenues: (1) via the initial data
  choice, which if constructed naively can excite logarithmic modes, (2) via 
  poorly chosen coordinates of the metric on the $u=$const hypersurfaces, and
  (3) via incomplete numerical
  cancellation in the equations, which possess nontrivial pole structure.
Points (1) and (2) arise from the use of the asymptotically nonflat Bondi form of the
spacetime metric, Eq.~(\ref{eq:BondiMetric}). In that form, even mathematically faithful solutions to the
hypersurface equations for generic worldtube data possess logarithmic dependence.
These logarithmic terms would vanish in an asymptotically flat coordinate system,
so they are a \emph{pure gauge} contribution.

In Secs. \ref{sec:InitialData} and~\ref{sec:radinull-integration}, we explain
  our methods for minimizing the logarithmic
contributions in the characteristic evolution system.
As part of the discussion in Sec.~\ref{sec:InitialData}, we describe
the choice of initial data that
eliminates logarithmic dependence from the first hypersurface of the
characteristic evolution system, addressing point (1)
 above. In Sec.~\ref{sec:radinull-integration}, we describe
improvements
to the integration techniques that address point (3)  above. These methods reduce logarithmic
  dependence to the point where it is not noticeable in the tests presented
  here. However, the full remedy for point (2) requires careful
reexamination of the characteristic evolution equations and a set of
coordinate transformations for the evolution system that will be considered
for future development of spectral characteristic techniques, but is beyond
the scope of this paper.

\subsection{Initial data slice} \label{sec:InitialData}
The characteristic evolution equations require boundary data on two
boundaries: the worldtube (thick red curve in Fig.~\ref{fig:SpaceTime})
and an initial slice $u=u_0$ (thick blue curve in Fig.~\ref{fig:SpaceTime}).
Boundary values on the worldtube were treated in Sec.~\ref{sec:IBF}
above; here we discuss values on the initial slice.
Given the hierarchical nature of the evolution equations, the only piece of the
metric we need to specify on the initial slice is $J$, as we can compute all of
the other evolution quantities from $J$ using
Eqs.~(\ref{eq:betaLin})-(\ref{eq:LinearizedAlphabetSoupEqns}).
The main mathematical consideration for choosing $J$ for the initial slice is
ensuring the regularity of $J$ at $\mathscr{I}^+$; the main physical
consideration in typical applications is choosing a $J$ that corresponds to no
incoming radiation, either by a linearized approximation~\cite{Babiuc:2010ze},
or by matching to a linearized solution~\cite{Bishop:2011iu}.
Finally, there is the numerical consideration mentioned in
  Sec.~\ref{sec:SpectralRepresentability}
  that we wish to minimize the excitation of
pure-gauge logarithmic dependence and keep the initial data $C^\infty$ over the
numerical domain.

When choosing $J$ on the initial $u=u_0$ slice, we
  wish to match the worldtube data provided by the Cauchy code as closely as
possible.
The worldtube data that we take as input (see Sec.~\ref{sec:IBF})
consist of the full spacetime metric and its first radial and
time derivatives, which are
sufficient to constrain the value of $J$ and the
value of $\partial_r J$ on the worldtube.
By careful analysis of the characteristic evolution
 equations,
one can show that
the initial $u=u_0$ hypersurface is free of logarithmic dependence if~\cite{TamburinoWinicour1966}
$\partial_\ell^2 J -J \left( (\partial_\ell K)^2 - \partial_\ell J
\partial_\ell \bar J\right) = 0$ at $\mathscr{I}^+$.
This condition is satisfied by the simpler conditions $J=J_{,\ell\ell}=0$ at
$\mathscr{I}^+$, so we construct an initial $J$ that satisfies
$J=J_{,\ell\ell}=0$ at $\mathscr{I}^+$ and matches the worldtube data.
This construction is consistent with the input Cauchy data in the overlap region of
(Fig.~\ref{fig:SpaceTime}) to linear order in a radial expansion.

Our initial choice of $J$, determined by the functions $J|_{\Gamma}$ and
$\partial_r J |_{\Gamma}$, is
\begin{align} \label{eq:J0initslice}
  J_{\text{initial}}
  =\,& \frac{R}{2r} \left(3 J|_{\Gamma} + R \partial_r J |_{\Gamma}\right)
    -\frac{R^3}{2 r^3} \left(J|_{\Gamma} + R \partial_r J|_{\Gamma}\right)\notag\\
  =\,& \frac{R}{2} \left(\frac{1}{\rho} - 1\right) \left(3 J|_{\Gamma} + R \partial_r J |_{\Gamma}\right)\notag\\
   &-\frac{R^3}{2}\left(\frac{1}{\rho} - 1\right)^3 \left(J|_{\Gamma} + R \partial_r J|_{\Gamma}\right).
\end{align}

\subsection{Radinull Integration}
\label{sec:radinull-integration}

  The characteristic equations Eqs.~(\ref{eq:betaLin})-(\ref{eq:LinearizedAlphabetSoupEqns}) can be solved in sequence by integration in $r$ from the worldtube to
  $\mathscr{I}^+$.  We use a numerical radinull grid in the
  compactified variable $\rho$, and we reexpress the characteristic equations
  in terms of $\rho$ derivatives; see Eqs.~(\ref{eq:betaEvo})-(\ref{eq:HEvo}).
  The grid points in $\rho$ are chosen at Chebyshev-Gauss-Lobatto quadrature
  points. 
The radinull equations for
$\beta_{,\rho}$ and $U_{,\rho}$ [Eqs.~(\ref{eq:betaEvo}) and (\ref{eq:UEvo})] both lend
themselves to straightforward
  Chebyshev-Gauss-Lobatto quadrature. Starting at the inner boundary values of
$\beta_{|\Gamma}$ [Eq.~(\ref{eq:betaIB})] and $U_{|\Gamma}$ [Eq.~(\ref{eq:UWIB})],
these evolution variables are integrated out to $\mathscr{I}^+$.

A quick examination of the radinull equations for the evolution quantities
$Q_{,\rho}, W_{,\rho},$ and $H_{,\rho}$ [Eqs.
~(\ref{eq:QEvo}), (\ref{eq:WEvo}), and (\ref{eq:HEvo})] reveals powers of
$(\rho - 1)$ in denominators, so regularity at $\mathscr{I}^+$($\rho = 1$) is
not guaranteed by the form of the equations.
A previous version of this same spectral characteristic evolution method~\cite{Handmer:2014} utilized
integration by parts in order to rewrite the equations in a form without poles,
allowing them to be integrated directly via Chebyshev-Gauss-Lobatto quadrature.
However, integration by parts introduced logarithmic terms like $\log(1-\rho)$
which canceled analytically in the final results of gauge invariants such as the
Bondi news, but which were not well represented by a Chebyshev-Gauss-Lobatto
spectral expansion in $\rho$.
These logarithmic terms spoiled exponential convergence and led to a large noise
floor, limiting the accuracy of the method.
We choose an alternative approach here.

The evolution equation for $Q$, Eq.~(\ref{eq:QEvo}), can be written in the form
\begin{align}
(r^2Q)_{,\rho} = \frac{Q_C}{(1-\rho)^2} + \frac{Q_D}{(1-\rho)^3},
\label{eq:QEvoRho}
\end{align}
where $Q_C$ corresponds to the $1/(1-\rho)^2$ term and $Q_D$ is
the $1/(1-\rho)^3$
term in Eq.~(\ref{eq:QEvo}), and all factors of $(1-\rho)$ in
  denominators have been written explicitly.

To better characterize the asymptotic behavior of this equation, we
rewrite the system in terms of
the inverse radinull coordinate $x = R/r = 1/\rho - 1$.
Then
Eq.~(\ref{eq:QEvoRho}) becomes
\begin{align}
\left(\frac{Q}{x^2}\right)_{,x} =
\frac{C}{x^2} + \frac{D}{x^3},
\label{eq:QEvoEll}
\end{align}
where
\begin{align}
C &= -\frac{Q_C+Q_D}{R^2}, \\
D &= -\frac{Q_D}{R^2}.
\end{align}

We know the right-hand side quantities $C$ and $D$ are regular at $x=0$, and we
seek a solution $Q$ that is also regular there. So we introduce new variables,
motivated by Taylor series expansions of $Q$, $C$, and $D$ about $\mathscr{I}^+$
($x=0$),
\begin{align}
\mathcal{Q} &= Q - Q_{|\mathscr{I}^+} - x Q_{,x|\mathscr{I}^+}, \\ 
\mathcal{C} &= C - C_{|\mathscr{I}^+} - x C_{,x|\mathscr{I}^+}, \\
\mathcal{D} &= D - D_{|\mathscr{I}^+} - x D_{,x|\mathscr{I}^+} - \frac{x^2}{2}D_{,xx|\mathscr{I}^+}.
\end{align}
Thus, by construction, $\mathcal{Q}$ and $\mathcal{C}$ are both guaranteed to
behave like $x^2$ near $x=0$ while $\mathcal{D}$ behaves as $x^3$.
Substituting these variables into Eq.~(\ref{eq:QEvoEll}) and gathering similar
terms, we find the differential equation
\begin{align}
\left(\frac{\mathcal{Q}}{x^2}\right)_{,x} &=
\frac{\mathcal{C}}{x^2} + \frac{\mathcal{D}}{x^3}
+ \frac{2C_{,x|\mathscr{I}^+}+D_{,xx|\mathscr{I}^+}}{2x} \nonumber\\
&+ \frac{Q_{,x|\mathscr{I}^+}+C_{|\mathscr{I}^+}+D_{,x|\mathscr{I}^+}}{x^2}
+ \frac{2Q_{|\mathscr{I}^+}+D_{|\mathscr{I}^+}}{x^3}.
\label{eq:ExpandedQEvox}
\end{align}
Because of how we have defined $\mathcal{Q}$, $\mathcal{C}$, and $\mathcal{D}$,
any potential singularity issues are confined to the last three terms.
To satisfy Eq.~(\ref{eq:ExpandedQEvox}) for all $x$, the numerators of each of
these terms
must identically vanish, providing constraints and boundary
conditions on the asymptotic values of $Q$, $C$, and $D$,
\begin{align}
Q_{|\mathscr{I}^+} &= -\frac{D_{|\mathscr{I}^+}}{2}, \\
Q_{,x|\mathscr{I}^+} &= -C_{|\mathscr{I}^+} - D_{,x|\mathscr{I}^+}, \\
0 &= -C_{,x|\mathscr{I}^+} - \frac{1}{2}D_{,xx|\mathscr{I}^+}.
\label{eq:RegularityQ}
\end{align}
The last equation, Eq.~(\ref{eq:RegularityQ}), is a regularity condition on $C$ and $D$.
If satisfied, it ensures no logarithmic dependence in the solution to the $Q$
equation.
A careful analysis of the differential equations, which will be presented in
complete detail in future work, shows that the leading violation of Eq.
~(\ref{eq:RegularityQ}) is
$\propto \bar{\eth} \partial_x^2 J |_{\mathscr{I}^+}$, and that Eq.
~(\ref{eq:RegularityQ}) is entirely satisfied if $J=0$ and $J_{,xx}$ = 0 at
$\mathscr{I}^+$.
The leading violation of the conditions on $J$ can be determined through further
analysis to have the leading contribution of
$U (\partial_x J)^2 |_{\mathscr{I}^+}$.
These pure-gauge regularity violations are important to note for precision
studies and for unusual regimes for characteristic evolution,
but for the practical evolutions, the scales we observe do not typically exceed
 $U\sim 10^{-6}$, $J \sim 10^{-3}$.
So, even for long evolutions, the logarithmic dependence does not grow to a
significant fraction of the main contribution.

We now integrate the equation
\begin{align}
\left(\frac{\mathcal{Q}}{x^2}\right)_{,x} &=
\frac{\mathcal{C}}{x^2} + \frac{\mathcal{D}}{x^3}
\end{align}
with inner boundary value
\begin{align}
  \mathcal{Q}_{|\Gamma} = Q_{|\Gamma} + \frac{D_{|\mathscr{I}^+}}{2} + \left(C_{|\mathscr{I}^+} + D_{,x|\mathscr{I}^+}\right)
\end{align}
to obtain $\mathcal{Q}$ at all radinull points. Then we reconstruct $Q$ by
adding back in its asymptotic values,
\begin{align}
Q = \mathcal{Q} - \frac{D_{|\mathscr{I}^+}}{2} - x\left(C_{|\mathscr{I}^+} + D_{,x|\mathscr{I}^+}\right).
\label{eq:QInt}
\end{align}
Because the equation for $Q$ does not mix the real and imaginary parts of $Q$,
we follow~\cite{Handmer:2014} and solve for real and imaginary parts of $Q$
separately.

Examining the evolution equation for $W$, Eq.~(\ref{eq:WEvo}), we
recognize that it has the same form as the equation for
$Q$, Eq.~(\ref{eq:QEvo}). Therefore, in
order to solve for $W$, we use the same procedure as we do for $Q$, following
from Eq.~(\ref{eq:QEvoRho}) through Eq.~(\ref{eq:QInt}) but replacing all of the
quantities specific to $Q$ with their $W$ equivalents.

The radinull equation for $H$, Eq.~(\ref{eq:HEvo}) can be written as
\begin{align}
(rH)_{,\rho} - \frac{rJ}{2}(H\bar T + \bar HT) = H_A + \frac{H_B}{1-\rho} + \frac{H_C}{(1-\rho)^2},
\label{eq:HEvoRho}
\end{align}
where $H_B = \Sigma_i H_{Bi}$.
The form of this equation is very similar to that of Eq.~(\ref{eq:QEvoRho}) that
governs the $Q$ (and $W$) radinull evolution. However, there is now the
additional complication that $H_{,\rho}$ has a term proportional to not
only $H$, but also to $\bar H$. This couples the real and imaginary parts of the
equation.

The previous version of this code employed the Magnus expansion in order to
handle this difficulty~\cite{Handmer:2014}. While the Magnus expansion might be
useful for systems where the terms in its expansion are rapidly shrinking, there
is no guarantee that will hold in general. Instead, we will write the system as
a matrix differential equation, expressing $H$ (and $H_A, H_B,$ and $H_C$) as
column vectors such as
\begin{align}
H = \left(
  \begin{array}{cc}
    \Re(H)\cr
    \Im(H)\cr
  \end{array}\right),
\end{align}
and defining the quantity $M$ as
\begin{align}
M \equiv \left(
  \begin{array}{cc}
    \Re(J)\Re(T) & \Re(J)\Im(T) \cr
    \Im(J)\Re(T) & \Im(J)\Im(T) \cr
  \end{array}\right),
\label{eq:MMatrix}
\end{align}
so that $MH$ here represents matrix multiplication. Then Eq.(\ref{eq:HEvoRho})
becomes the matrix equation,
\begin{align}
(rH)_{,\rho} - rMH = H_A + \frac{H_B}{1-\rho} + \frac{H_C}{(1-\rho)^2},
\label{eq:HEvoRhoMatrix}
\end{align}

As before, we convert from $\rho$ into the inverse radinull coordinate
$x=R/r=1/\rho - 1$ to better characterize its behavior near $\mathscr{I}^+$,
\begin{align}
\left(\frac{H}{x}\right)_{,x} + \mathcal{M}\frac{H}{x} =
    A + \frac{B}{x} + \frac{C}{x^2}
\label{eq:HEvoEll}
\end{align}
where
\begin{align}
\label{eq:McalMatrix}
\mathcal{M} =& \frac{M}{(1+x)^2}, \\
A =& -\frac{H_A}{R(1+x)^2}, \\
B =& -\frac{H_B}{R(1+x)}, \\
C =& -\frac{H_C}{R}.
\end{align}
As we did
with the $Q$ equation, we shall introduce one final set of variables,
motivated by Taylor series expansions of $H, B,$ and $C$ about $x=0$:
\begin{align}
\mathcal{H} =& H - H_{|\mathscr{I}^+}, \\
\mathcal{B} =& B - B_{|\mathscr{I}^+} - \mathcal{M}H_{|\mathscr{I}^+} + \mathcal{M}_{|\mathscr{I}^+}H_{|\mathscr{I}^+}, \\
\mathcal{C} =& C - C_{|\mathscr{I}^+} - x C_{,x|\mathscr{I}^+}.
\end{align}
Once again, these variables are constructed so that $\mathcal{H}$ and
$\mathcal{B}$ behave as $x$ and $\mathcal{C}$ behaves as $x^2$ in a
neighborhood about $x=0$. Substituting these into Eq.~(\ref{eq:HEvoEll}), we
get
\begin{align}
\left(\frac{\mathcal{H}}{x}\right)_{,x} + \mathcal{M}\frac{\mathcal{H}}{x} =&
    A + \frac{\mathcal{B}}{x} + \frac{\mathcal{C}}{x^2}
    + \frac{H_{|\mathscr{I}^+}+C_{|\mathscr{I}^+}}{x^2} \nonumber\\
    +& \frac{B_{|\mathscr{I}^+} + C_{,x|\mathscr{I}^+} - \mathcal{M}_{|\mathscr{I}^+}H_{|\mathscr{I}^+}}{x}
\end{align}
As before, the numerators of the last two terms must vanish,
  which gives us a boundary condition on $H$ at $\mathscr{I}^+$,
\begin{align}
H_{|\mathscr{I}^+} = -C_{|\mathscr{I}^+},
\end{align}
and a boundary constraint on $B$, $C$, and $\mathcal{M}$,
\begin{align}
0 = B_{|\mathscr{I}^+} + C_{,x|\mathscr{I}^+} + \mathcal{M}_{|\mathscr{I}^+}C_{|\mathscr{I}^+}.
\label{eq:HBdryConstraint}
\end{align}
The last constraint is a regularity condition that is
guaranteed to be satisfied provided the input spin-weighted scalars
  $\beta, Q, U,$ and $W$ themselves are regular~\cite{TamburinoWinicour1966}.
  Of course, the small violation that arises from the $Q$ and $W$ equations
  will lead to a similarly small violation in the regularity of $H$.
  In principle, a carefully chosen coordinate transformation could fully
  address all of these small violations.

We then integrate the equation
\begin{align}
\left(\frac{\mathcal{H}}{x}\right)_{,x} + \mathcal{M}\frac{\mathcal{H}}{x} =&
    A + \frac{\mathcal{B}}{x} + \frac{\mathcal{C}}{x^2}
\end{align}
from the worldtube to $\mathscr{I}^+$, with boundary value
$\mathcal{H}_{|\Gamma} = H_{|\Gamma} + C_{|\mathscr{I}^+}$, to obtain
$\mathcal{H}$ on the entire null slice. We reconstruct $H$ by computing
\begin{align}
H = \mathcal{H} - C_{|\mathscr{I}^+}.
\end{align}

To help ensure the stability of the system, we perform spectral filtering for
each of the evolution quantities
$J$, $\beta$, $Q$, $U$, $W$, and $H$ after every time we compute them,
  similar to~\cite{Handmer:2014}. For the angular filtering,
  we set to 0 the highest two
  $\ell$-modes in the spectral decomposition on each shell of constant $\rho$.
  Thus, resolving the
system up through $\ell_{\text{max}}$ modes requires storing and evolving the
evolution quantities in the volume up through $\ell=\ell_{\text{max}}+2$ modes.
We filter along the radinull direction by taking the spectral expansion of the evolution
quantities along each null ray and scaling the $i$th coefficient by
\begin{align}
e^{-108\left(i/(n_{\rho}-1)\right)^{16}},
\end{align}
where $n_\rho$ is the number of radinull points.
This is a fairly stringent filter. Future work may be able to retain
more mode content by exploring the precise needs of the filter to avoid
aliasing effects in a range of practical simulation data.

\begin{figure}[t!]
\includegraphics[width=.97\columnwidth]{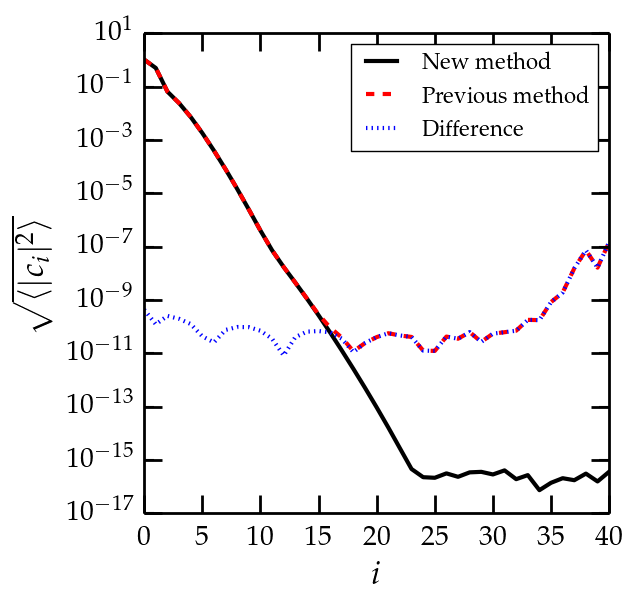}
\caption{
The angle-averaged value of the radinull spectral coefficients of $H$
after integrating Eq.~(\ref{eq:HEvoEll}) for the test system given
in Eqs.~(\ref{eq:testA})-(\ref{eq:testGamma}) for both the 
new method of integration described here and the previous method
introduced in~\cite{Handmer:2014}.
\label{fig:RadIntH}
}
\end{figure}

To demonstrate the
      improvement afforded by our new method of treating the radinull
    integration,
  we test the new
    method of integrating Eq.~(\ref{eq:HEvoEll}) versus the previous
  method introduced in~\cite{Handmer:2014} on an analytic test case.
    Consider Eq.~(\ref{eq:HEvoEll}) with
\begin{align}
  A =& \frac{.94\sin x-1.53\cos x}{R(1+x)^2}\ _2Y_{31}\ _0Y_{2-2}\ _0Y_{10},
  \label{eq:testA}\\
B =& -\frac{x\sin{x}}{R}\left(_2Y_{3-1}+_3Y_{4-3}\ _1Y_{22}\right) \nonumber\\
   & + \left(_0Y_{00}-1\right){}_2Y_{22}|_2Y_{22}|^2 \\
C =& (1-\cos x)\left(_2Y_{3-1}+_1Y_{2-2}\ _1Y_{30}\right) \nonumber\\
   & + (\sin x+\cos x){}_2Y_{22},
\end{align}
and with $\mathcal{M}$ defined
by Eqs.~(\ref{eq:MMatrix}),~(\ref{eq:McalMatrix}),
  and~(\ref{eq:TOrGamma}) with
\begin{align}
J =& _2Y_{22}, \\
T =& {}_2Y_{22}\ {}_0Y_{00}. \label{eq:testGamma}
\end{align}
For this test case, we set $R=2.94$ and
we resolve the computational domain 
through $L=10$ and $n_\rho=41$. This test case is
not necessarily physical,
but satisfies the boundary constraint given in Eq.~(\ref{eq:HBdryConstraint}).
We integrate Eq.~(\ref{eq:HEvoEll}) from an inner boundary value of
$H_{|\Gamma}=0$ to $x=0$, obtaining $H$
as a function of $x$,$\theta$, and $\phi$, or equivalently, obtaining
the radial spectral coefficients of $H$, $c_i(\theta_j,\phi_k)$ at each
angular collocation point $(\theta_j,\phi_k)$.
To reduce the size of the dataset, we average these coefficients over
the sphere according to
\begin{align}
  \sqrt{\langle|c_i|\rangle} =&
  \sum_{j,k}c_i(\theta_j,\phi_k)
  \bar{c_i}(\theta_j,\phi_i)\sin{\theta_j}\frac{2\pi^2}{n_\theta n_\phi},
\end{align}
and we plot these angle-averaged coefficients in Fig.~\ref{fig:RadIntH}.

Because the test case
    satisfies the regularity conditions,
  we expect that with sufficient resolution,
  an accurate integration
  scheme would be capable of resolving
  the solution to
  numerical roundoff. From Fig.~\ref{fig:RadIntH} we see
    that our current method demonstrates this behavior. However,
  the radial modes of the previous method from~\cite{Handmer:2014}
  flattens out 
  about 6 orders of magnitude
  larger, because the logarithmic terms
  are not properly
  represented via our chosen spectral decomposition.

\subsection{Time evolution}\label{subsec:TimeEvolution}

To evolve $J$ forward in time, we integrate
\begin{align}
J_{,u|\rho,x^A=\text{const}} = \Phi
\label{eq:EvolutionEquationJ}
\end{align}
at each radinull point using the method of lines. This is done using an ordinary differential equation (ODE)
integrator, integrating forward in $u$, with a supplied right-hand side $\Phi$.
Here $\Phi$ is computed using
\begin{align}
\Phi = H + \rho(1-\rho)\frac{R_{,\bar u}}{R}J_{,\rho},
\label{eq:PhiFromH}
\end{align}
where $R_{,\bar u}$ is the derivative of the surface area coordinate $r$ along the worldtube
given by Eq.~(\ref{eq:WTBondiRDerivU}) and where $H$ is the result of the
radinull integration, Eq.~(\ref{eq:LinearizedAlphabetSoupEqns}), accomplished
using the method in Sec.~\ref{sec:radinull-integration}.

The time integration of $J$ [Eq.~(\ref{eq:EvolutionEquationJ})] uses a fifth order
Dormand-Prince ODE solver with adaptive time stepping~\cite{Press2007}, and a
default relative error tolerance of $10^{-8}$ except where otherwise noted.
The step sizes are limited entirely by the error measure and is
independent of the time steps of the Cauchy evolution used to generate the
worldtube. The time evolution is also done in tandem with the evolution of the
inertial coordinates [Eq.~(\ref{eq:InerCoord}),
and of the conformal factor (Eq.~(\ref{eq:ScriDomegadu})] from $\mathscr{I}^+$ extraction,
as described below. 

\section{$\mathscr{I}^+$ Extraction}\label{sec:SE}

Once the characteristic equations have been solved in the volume so that the
metric variables of the Bondi-Sachs form
  Eq.~(\ref{eq:BondiMetric}) are known on $\mathscr{I}^+$, the gravitational
waveform can be computed.
This involves two steps.
The first step is computing the Bondi news function at $\mathscr{I}^+$ from the
metric variables there.
The second step involves transforming the news to a freely falling coordinate
system at $\mathscr{I}^+$; this removes all remaining gauge freedom up to a BMS
transformation.
These steps are described below.

\subsection{News function}
\label{sec:news-function}

The metric in Bondi-Sachs form given in
Eq.~(\ref{eq:BondiMetric}) is divergent at $\mathscr{I}^+$ where
$r\rightarrow\infty$, so we work with a conformally rescaled Bondi metric,
$\hat g_{\mu\nu}=\ell^2g_{\mu\nu}$, where $\ell=1/r$, that is finite at
$r\rightarrow\infty$.
Expressing this metric in the coordinate system
$\hat x^{\hat\alpha} = (u,\ell,x^A)$, it takes the form~\cite{Bishop:1997ik}
\begin{align}
\hat g_{\hat\mu\hat\nu} =& -\left(e^{2\beta}(\ell^2+\ell W) - h_{AB}U^AU^B\right)du^2
    \nonumber\\
    +& 2e^{2\beta}dud\ell - 2h_{AB}U^Bdudx^A \nonumber\\
    +& h_{AB}dx^Adx^B.
\label{eq:BondiMetricCompact}
\end{align}
Here $h_{AB}$, $\beta$, $W$, and $U^A$ are the same quantities that appear in
Eq.~(\ref{eq:BondiMetric}).

To facilitate the computation of the news function, we construct an additional
conformal metric
\begin{align}
\tilde g_{\hat\mu\hat\nu}&=\omega^2\hat g_{\hat\mu\hat\nu},
\label{eq:BondiMetricConformal}
\end{align}
that is asymptotically Minkowski at $\mathscr{I}^+$. The conformal
factor $\omega$ is chosen so that the angular part of $\tilde g_{\mu\nu}$ is a
unit sphere metric~\cite{TamburinoWinicour1966},
\begin{align}
q_{AB} = \omega^2h_{AB|\mathscr{I}^+}.
\label{eq:ConformalOmega}
\end{align}
In terms of the original metric,
\begin{align}
\tilde g_{\hat\mu\hat\nu} =& \Omega g_{\hat\mu\hat\nu},\\
\Omega =& \omega\ell.
\label{eq:OmegaDefinition}
\end{align}

On a given constant $u$ slice, $\omega$ can be computed by solving an elliptic
equation related to the two-dimensional (2D) curvature scalar,
\begin{align}
\mathcal{R}=2\left(\omega^2 + h^{AB}_{|\mathscr{I}^+}D_AD_B\ln\omega\right),
\label{eq:OmegaEllipticEq}
\end{align}
where $D_A$ is the covariant derivative associated with
$h^{AB}_{|\mathscr{I}^+}$.
Equation~(\ref{eq:OmegaEllipticEq}) has the effect of setting the asymptotic 2D
  curvature in the conformally rescaled metric to be $2$, which is the curvature
  of the unit sphere.
Expanding out the covariant derivatives yields~\cite{Bishop:1997ik},
\begin{widetext}
\begin{align}
h^{AB}_{|\mathscr{I}^+}D_AD_B\ln\omega =& \frac{1}{4}\Big(
    - 2\eth^2\ln\omega\bar J - 2\bar\eth^2\ln\omega J + 4\bar\eth\eth\ln\omega K
    - \eth\ln\omega\eth J\bar J^2 - \eth\ln\omega\eth\bar JJ\bar J
    - 2\eth\ln\omega\eth\bar J \nonumber\\
   +& 2\eth\ln\omega\eth K\bar JK + \eth\ln\omega\bar\eth J\bar JK
    + \eth\ln\omega\bar\eth\bar JJK - 2\eth\ln\omega\bar\eth KJ\bar J
    + \eth J\bar\eth\ln\omega\bar JK +\eth\bar J\bar\eth\ln\omega JK \nonumber\\
   -& 2\eth K\bar\eth\ln\omega J\bar J - \bar\eth\ln\omega\bar\eth JJ\bar J
    - 2\bar\eth\ln\omega\bar\eth J - \bar\eth\ln\omega\bar\eth\bar JJ^2
    + 2\bar\eth\ln\omega\bar\eth KJK \Big).
\label{eq:OmegaConstr}
\end{align}
\end{widetext}
Equation~(\ref{eq:OmegaEllipticEq})
could in principle be used to solve for $\omega$ at each slice
of constant $u$.  However, we instead solve this equation for $\omega$
only on the initial slice, where the equation simplifies significantly (see below), and
then we construct an evolution equation for $\omega$
and we evolve $\omega$ as a function of $u$.  Note that when evolving
$\omega$, one could use Eq.~(\ref{eq:OmegaEllipticEq}) as a check to monitor
the error in $\omega$; however we do not yet do so.

On the initial slice, Eqs.~(\ref{eq:OmegaEllipticEq}) and~(\ref{eq:OmegaConstr})
simplify considerably;
we have set $J_{|\mathscr{I}^+}=0$ (see Eq.~(\ref{eq:J0initslice})), so
Eq.~(\ref{eq:OmegaConstr}) implies that $h^{AB}_{|\mathscr{I}^+}D_AD_B\ln\omega
= 4\bar\eth\eth\ln\omega$ and Eq.~(\ref{eq:Rcurve}) implies that $\mathcal{R} =
2$, reducing Eq.~(\ref{eq:OmegaEllipticEq}) to
$1=\omega^2+\bar\eth\eth\ln\omega$. This has the trivial solution of $\omega=1$.

The null generators at $\mathscr{I}^+$ are defined
as~\cite{Bishop:1997ik},
\begin{align}
\tilde n^{\hat\mu} = &\tilde g^{\hat\mu\hat\nu}\nabla_{\hat\nu}\Omega_{|\mathscr{I}^+},
    \label{eq:ConformalGenerator} \\
\hat n^{\hat\mu} = &\hat g^{\hat\mu\hat\nu}\nabla_{\hat\nu}\ell_{|\mathscr{I}^+} =\hat g^{\hat\mu\ell},
    \label{eq:CompactifiedGenerator}
\end{align}
so that
\begin{align}
\tilde n^{\hat\mu} =\omega^{-1}\hat n^{\hat\mu}.
\end{align}
where the covariant derivative $\nabla_{\hat\nu}$ is associated with the Bondi
metric, $g_{\hat\mu\hat\nu}$. Derivation for evolution of the conformal factor on
$\mathscr{I}^+$ in the frame of the compactified metric, is given in
Ref~\cite{Bishop:1997ik}, and can be computed by
\begin{align}
2\hat n^{\hat\mu}\nabla_{\hat\mu}\ln\omega = -e^{-2\beta}W_{|\mathscr{I}^+}.
\label{eq:ScriDomegadu}
\end{align}

Reference~\cite{Bishop:1997ik} derived the formula for the news function in the
conformal metric with the evolution coordinates, with a sign error corrected
in~\cite{BabiucEtAl2008} (Ref~\cite{Bishop:1997ik}
chose their convention to agree with Bondi's
original expression
in the axisymmetric case~\cite{Bondi1962}). Here we have
factored the $s_i$ slightly differently than they
did,
\begin{widetext}
\begin{align}
N =& \frac{1}{16\omega A(K+1)}\left(4s_1 + 2s_2
    - \left(\eth\bar U+\bar\eth U\right)s_3 - \frac{8}{\omega^2}s_4
    + \frac{2}{\omega}s_5\right),
    \label{eq:NewsFunction}\\
A =& \omega e^{2\beta}, \label{eq:NewsA}\\
s_1 =& J^2\bar H_{,\ell} + J\bar JH_{,\ell}
    + 2(K+1)\left(H_{,\ell} - JK_{,u\ell}\right), \\
s_2 =& \eth J_{,\ell}J\bar J\bar U + \eth\bar J_{,\ell}J^2\bar U + 2\eth UJ\bar JK_{,\ell}
    + 2\eth\bar UJ\bar JJ_{,\ell} + \bar\eth J_{,\ell}J\bar JU
    + \bar\eth\bar J_{,\ell}J^2U + 2\bar\eth UJ^2\bar J_{,\ell}
    + 2\bar\eth\bar UJ^2K_{,\ell} \nonumber\\
    +& (K+1)\Big(2\eth J_{\ell}\bar U - 2\eth K_{,\ell}J\bar U - 2\eth UJ\bar J_{,\ell}
    + 4\eth UK_{,\ell} - 2\eth\bar UJK_{,\ell} + 4\eth\bar UJ_{,\ell}
    + 2\bar\eth J_{,\ell}U - 2\bar\eth K_{,\ell}JU \nonumber\\
    -& 2\bar\eth UJK_{,\ell}- 2\bar\eth\bar UJJ_{,\ell}\Big), \\
s_3 =& J^2\bar J_{,\ell} + J\bar JJ_{,\ell} + 2(K+1)(J_{,\ell} - JK_{,\ell}), \\
s_4 =& \eth A\eth\omega J\bar J + \bar\eth A\bar\eth\omega J^2
    + (K+1)\Big(2\eth A\eth\omega - \eth A\bar\eth\omega J
    - \bar\eth A\eth\omega J\Big), \\
s_5 =& 2\eth^2AJ\bar J + 2\bar\eth^2AJ^2 + \eth A\eth JJ\bar J^2
    + \eth A\eth\bar JJ^2\bar J  - \eth A\bar\eth JJ\bar JK
    - \eth A\bar\eth\bar JJ^2K + 2\eth A\bar\eth KJ^2\bar J \nonumber\\
    +& 2\bar\eth A\eth KJ^2\bar J
    + \bar\eth A\bar\eth JJ^2\bar J + \bar\eth A\bar\eth\bar JJ^3
    - 2\bar\eth A\bar\eth KJ^2K \nonumber\\
    +& (K+1)\Big(4\eth^2A - 4\bar\eth\eth AJ + 2\eth A\eth J\bar J
    + 2\eth A\eth\bar JJ - 4\eth A\eth K + 2\eth A\bar\eth J - 2\bar\eth A\eth J
    + 4\bar\eth A\eth KJ\Big) \nonumber\\
    +& (K+2)\Big(-2\eth A\eth KJ\bar J -\bar\eth A\eth JJ\bar J
    - \bar\eth A\eth\bar JJ^2\Big).
\end{align}
\end{widetext}

The news as defined in Eq.~(\ref{eq:NewsFunction}) has spin weight $+2$.
However, the usual convention for gravitational radiation is to work with
quantities with spin weight $-2$. Furthermore, the news $N$ has the opposite
sign as the usual convention. 
To relate this news function to the gravitational wave strain defined
using the following convention: given a radially outward propagating metric
perturbation from Minkowski, $h_{\tilde\mu\tilde\nu} = g_{\tilde\mu\tilde\nu} -
\eta_{\tilde\mu\tilde\nu}$ and polarizations given by $h_{+} =
(h_{\tilde\theta\tilde\theta} + h_{\tilde\phi\tilde\phi})/2$ and $h_{\times} =
h_{\tilde\theta\tilde\phi}$, the strain is given by
\begin{align}
h = h_{+} - ih_{\times}.
\end{align}
Then the news is related to the strain by,
\begin{align}
\partial_{\tilde u}h = 2\overline{N}.
\end{align}

\subsection{Inertial coordinates}\label{subsec:InerCoords}

Once the news function is computed according to
  Sec.~\ref{sec:news-function}, it is known as a function of
  coordinates $(u,x^A)$ on $\mathscr{I}^+$.  Recall that these
  coordinates are chosen so that  $u=\breve t$ and
  $x^A = \breve x^{\breve A}$ on the worldtube,
  where $(\breve t,\breve x^{\breve A})$
  are the time and angular coordinates of the interior Cauchy
  evolution.  Therefore, the news as computed above depends on the
  choice of Cauchy coordinates.

In this section, we transform the news to a new inertial
  coordinate system $(\tilde u,\tilde x^{\tilde A})$ on
  $\mathscr{I}^+$, where curves of constant $\tilde x^{\tilde A}$
  correspond to worldlines of free-falling observers
  (because we are working on $\mathscr{I}^+$, we can suppress the
  radinull coordinate). This removes
  the remaining gauge freedom in the news, up to a choice of
  free-falling observers (or in other words up to a BMS transformation).

On the initial slice, we choose $\tilde u=u$ and $\tilde x^{\tilde A}=x^A$.
These inertial coordinates then evolve along the $\mathscr{I}^+$
generators~\cite{Bishop:1997ik},
\begin{align}
\hat n^\mu\partial_\mu\tilde u = \omega, \\
\hat n^\mu\partial_\mu\tilde x^{\tilde A} = 0,
\label{eq:InertialAngEvolution}
\end{align}
where the $\hat n^\mu$ are given by elements of the compactified metric
according to Eq.~(\ref{eq:CompactifiedGenerator}).

Since $\tilde x^{\tilde A}=(\tilde\theta,\tilde\phi)$ are not representable via
a spectral expansion in spherical harmonics, thus making them poor choices for
our numerics, we represent the inertial coordinates using a Cartesian basis
$\tilde x^{\tilde\i}=(\tilde x,\tilde y, \tilde z)$. We reexpand
Eq.~(\ref{eq:InertialAngEvolution}), using the transformations
\begin{align}
\frac{\partial\tilde\theta}{\partial x^\mu} &=
    \frac{1}{\tilde x^2+\tilde y^2}
    \left(-\tilde y\frac{\partial\tilde x}{\partial x^\mu}
    + \tilde x\frac{\partial\tilde y}{\partial x^\mu}\right),
    \label{eq:dthetadxyz} \\
\frac{\partial\tilde\phi}{\partial x^\mu} &=
    \frac{1}{\tilde r^2\sqrt{\tilde x^2+\tilde y^2}}
    \left(\tilde x\tilde z\frac{\partial\tilde x}{\partial x^\mu}
    + \tilde y\tilde z\frac{\partial\tilde y}{\partial x^\mu}
    - (\tilde x^2+\tilde y^2)\frac{\partial\tilde z}{\partial x^\mu}\right).
    \label{eq:dphidxyz}
\end{align}
Plugging those into Eq.~(\ref{eq:InertialAngEvolution}) yields the coupled equations
\begin{widetext}
\begin{align}
-\tilde y\frac{\partial\tilde x}{\partial u}
    + \tilde x\frac{\partial\tilde y}{\partial u} &=
    \frac{\hat n^{\hat A}}{\hat n^u}\left(-\tilde y\frac{\partial\tilde x}{\partial \hat{x}^{\hat A}}
    + \tilde x\frac{\partial\tilde y}{\partial \hat{x}^{\hat A}}\right),
    \label{eq:dTHETAdu} \\
\tilde x\tilde z\frac{\partial\tilde x}{\partial u}
    + \tilde y\tilde z\frac{\partial\tilde y}{\partial u}
    - (\tilde x^2+\tilde y^2)\frac{\partial\tilde z}{\partial u} &=
    \frac{\hat n^{\hat A}}{\hat n^u}\left(\tilde x\tilde z\frac{\partial\tilde x}{\partial \hat{x}^{\hat A}}
     +\tilde y\tilde z\frac{\partial\tilde y}{\partial \hat{x}^{\hat A}}
     -(\tilde x^2+\tilde y^2)\frac{\partial\tilde z}{\partial \hat{x}^{\hat A}}\right).
    \label{eq:dPHIdu}
\end{align}
\end{widetext}

By expanding the basis from two coordinates to three, we also need to introduce a
constraint which will force the $\tilde x^{\tilde\i}$ to remain on the unit sphere and
eliminate the extra degree of freedom, $\tilde r=\sqrt{\tilde x^2+\tilde
y^2+\tilde z^2}=1$. While this holds analytically, numerically $\tilde r$ will
shift away from 1 during the evolution, which makes it necessary to introduce a
constraint equation to the system of equations,
\begin{align}
\frac{\partial\tilde r}{\partial u} = \tilde x\frac{\partial\tilde x}{\partial u}
    + \tilde y\frac{\partial\tilde y}{\partial u} + \tilde z\frac{\partial\tilde z}{\partial u}
    = \tilde r C(\tilde r),
    \label{eq:ScriRConstraintEqn}
\end{align}
where $C(\tilde r)$ is a constraint term where $C(\tilde r=1)=0$. In our code,
$C(\tilde r)=-\kappa(\tilde r-1)$ for some positive parameter $\kappa$.

With these three equations,
Eqs.~(\ref{eq:dTHETAdu})-(\ref{eq:ScriRConstraintEqn}), we solve for
the three $\frac{\partial\tilde x^{\tilde\i}}{\partial u}$. After some manipulations and
massaging, we obtain the evolution equations for the Cartesian inertial
coordinates with respect to the characteristic coordinates,
  \begin{align}\label{eq:InerCoord}
    \frac{\partial \tilde{x}^i}{\partial u} =
    \frac{\tilde{x}^i}{\tilde{r}}C(\tilde{r})
    + \frac{1}{\tilde{r}^2}
    \left( -\tilde{x}^i\tilde{x}^j \delta_{jk} + \delta^{i}{}_k \tilde{r}^2 \right)
    \frac{\partial \tilde{x}^k}{\partial \hat{x}^{\hat{A}}}
    \frac{\hat{n}^{\hat{A}}}{\hat{n}^{u}}.
  \end{align}

Once we know $\tilde u(u,x^A)$, $\tilde x^{\tilde\i}(u,x^A)$, then obtaining the news
on this grid is a matter of interpolation. Our code does so in
two steps. First, each of the spatial coordinates, as well as the news function
is interpolated in time onto slices of constant $\tilde u$, so that we then have
both $\tilde x^{\tilde i}(\tilde u,x^A)$ and $N(\tilde u,x^A)=N(\tilde u,\tilde x^{\tilde\i})$,
using a cubic spline along each grid point on $\mathscr{I}^+$.

Then on each constant $\tilde u$ slice, we perform the spatial interpolation by
projecting the news function onto its spectral coefficients $c^{\ell m}$, using
the orthonormality of SWSHes from Eq.~(\ref{eq:SWSHOrthonormality}),
\begin{align}
c^{\ell m}(\tilde u) = \int_{S^2}N(\tilde u,\tilde x^{\tilde\i})
    \overline{^{2}Y^{\ell m}}(\tilde\theta,\tilde\phi)
    \sin\tilde\theta d\tilde\theta d\tilde\phi.
\end{align}
However, since we numerically evaluate news function on the noninertial
characteristic coordinates, we must instead do the integration over its area
elements, $\sin\theta d\theta d\phi$, so we convert the coordinates of this
expression, which introduces the determinant of a Jacobian,
\begin{align}
d\tilde\theta d\tilde\phi = 
    d\theta d\phi \left|\frac{\partial\tilde x^{\tilde A}}{\partial x^{A}}\right|.
\end{align}
Once again, because of the difficulties of representing angular coordinates
spectrally, we convert this expression from $\tilde\theta$ and $\tilde\phi$ to
$\tilde x^{\tilde\i}$. To facilitate our expansion to Cartesian coordinates, we introduce
a temporary radial coordinates $\tilde{\mathfrak{r}}$ and $\mathfrak{r}$ on the
unit sphere with $\tilde x^{\tilde A}=(\tilde{\mathfrak{r}}, \tilde\theta, \tilde\phi)$ and $x^A=(\mathfrak{r}, \theta, \phi)$ so that we can properly define the determinants (keeping in mind
$\tilde{\mathfrak{r}}$ and $\mathfrak{r}$ are analytically identical to 1 and
will disappear from the final expressions),
\begin{align}
\left|\frac{\partial\tilde x^{\tilde A}}{\partial x^{A}}\right| =&
    \left|\frac{\partial\tilde x^{\tilde A}}{\partial\tilde x^{\tilde\i}}\right|
    \left|\frac{\partial\tilde x^{\tilde\i}}{\partial x^{A}}\right| \nonumber\\
    =&\left(\frac{1}{\tilde{\mathfrak{r}}^2\sin\tilde\theta}\right)
    \left|\frac{\partial\tilde x^{\tilde\i}}{\partial x^{A}}\right|.
\end{align}

Plugging everything in yields the full expression,
\begin{align}
c^{\ell m}(\tilde u) = \int_{S^2}N(\tilde u,\tilde x^{\tilde\i})\overline{^{2}Y^{\ell m}}(\tilde\theta,\tilde\phi)
    \frac{1}{\sin\theta}\left|\frac{\partial\tilde x^{\tilde\i}}{\partial x^{A}}\right|
    \sin\theta d\theta d\phi.
\label{eq:InerCoordProjection}
\end{align}
Note that we have included a factor of $\sin\theta/\sin\theta$ which, while
analytically trivial, aids with the numerics of our code. Incorporating the
$\sin\theta$ in the numerator generates the proper spherical area element for
the integration, while we factor the $1/\sin\theta$ into the
$\frac{\partial}{\partial\phi}$ terms in the Jacobian, as numerically computed
spherical gradients return factors of
$\frac{1}{\sin\theta}\frac{\partial}{\partial\phi}$.

If the strain is similarly decomposed into spin weight $-2$ spherical
harmonic coefficients, $h_{\ell m}$, then they are related to the news
coefficients by,
\begin{align}
\partial_{\tilde u}h_{\ell m} = 2(-1)^{-m}\overline{c^{\ell -m}_{\text{CCE}}}.
\end{align}

One potential issue with Eq.~(\ref{eq:InerCoordProjection}) is the possibility that
there is a significant drift in the inertial coordinates relative to the code
coordinates. If there is a large systematic shift in the coordinates (for
example, if they all drift toward a single sky location), then there could be
regions on the unit sphere which are sparsely represented. Because spectral
methods of computing integrals often assume an optimal distribution of grid
points across the surface, this drift means there is a risk of underresolving
the computation Eq.~(\ref{eq:InerCoordProjection}), especially for high $\ell$
modes. To forestall this issue, we have taken to representing the $\mathscr{I}^+$
extraction portion at a significantly higher angular resolution from the rest of
our code. In particular, when we properly resolve the volume evolution up to
$\ell_{\text{max}}$ angular modes, we maintain a basis consisting of $2\ell_{\text{max}}$
angular modes for our $\mathscr{I}^+$ extraction code. Our properly resolved information
content is still no better than what is resolved in the volume evolution (i.e.
$\ell_{\text{max}}$), but this allows us to accurately project onto the inertial
coordinates with Eq.~(\ref{eq:InerCoordProjection}). Because the $\mathscr{I}^+$ extraction
portion of the code is only a 2D surface, this choice is an insignificant
contribution to the overall computational cost of our code.

While this coordinate evolution projects the news function on an inertial frame,
it is not a unique inertial frame. The class of inertial observers at
$\mathscr{I}^+$ are all related to each other by the group of BMS
transformations. Because our CCE inertial coordinates at $\mathscr{I}^+$
correspond to free-falling observers, the BMS frame remains constant throughout
the entire characteristic evolution. Thus, the BMS frame we use in our evolution
is frozen in entirely by our choice to identify our inertial coordinates with
the characteristic coordinates on our initial slice (i.e. $\tilde u=u$
and $\tilde x^{\tilde A}=x^A$). This choice is in some sense arbitrary, as it is
ultimately related to the coordinates provided on the worldtube by the Cauchy
evolution on that initial slice, and there are no guarantees of consistency between
CCE evolutions on different worldtubes even from the same Cauchy evolution.
However, development of a consistent treatment of handing the choice of BMS
frame is beyond the scope of this paper.

\subsection{Computational grid}

We use {\tt Spherepack} for most of the $\mathscr{I}^+$ Extraction, with the final projection
onto the inertial coordinates done using {\tt Spinsfast}. The time evolution of the
inertial coordinates, Eq.~(\ref{eq:InerCoord}), and
of the conformal factor, Eq.~(\ref{eq:ScriDomegadu}), is done in tandem with the
evolution of $J$, Eq.~(\ref{eq:PhiFromH}), in the volume extraction, using the same
routine (fifth order Dormand-Prince) and error tolerance as specified for that
evolution.

\section{Code Tests}\label{sec:CodeTests}

In order to showcase the accuracy, speed, and robustness of this spectral CCE
code, we perform a number of tests on the code. We have two linearized
solutions, a trivial analytic solution, and two fully nonlinear tests which
outline how well the code can remove purely coordinate effects from the news
output.

\subsection{Linearized analytic solution}\label{subsec:LinAnaSol}

\begin{figure}[t!]
\includegraphics[width=.97\columnwidth]{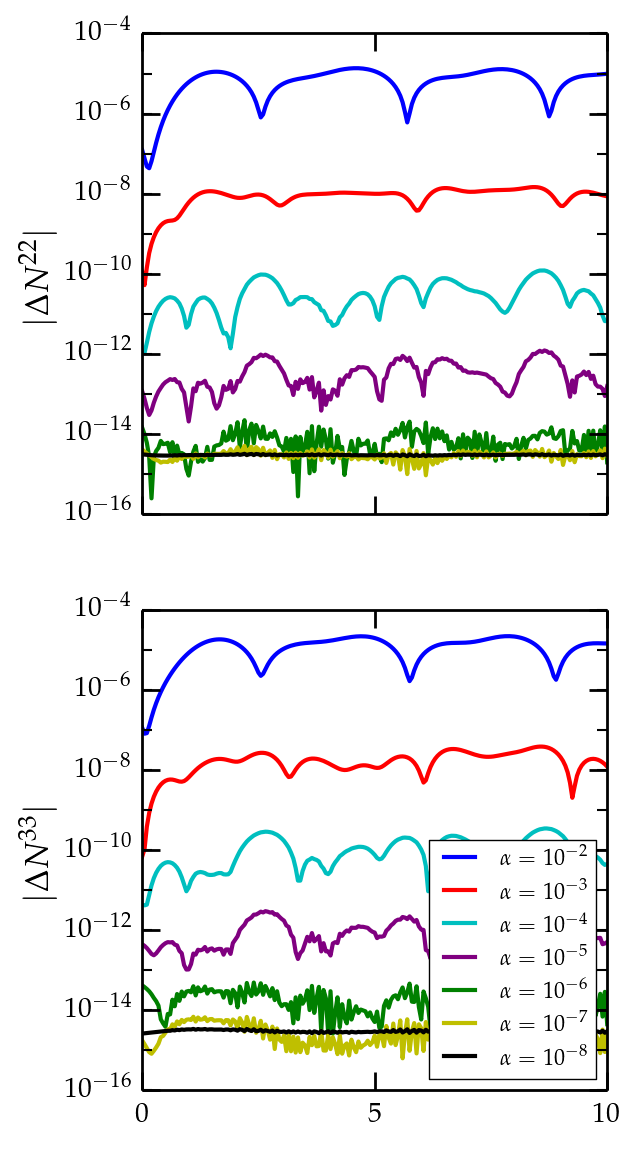}
\caption{
The difference between the numerically evolved news function and the analytic
solution for the linearized analytic test of Sec.~\ref{subsec:LinAnaSol}, for
various amplitudes of the linear perturbation $\alpha$. The $(2,2)$ mode is on the
left and the $(3,3)$ mode on the right. We expect differences of order $\alpha^2$
because we evolve the nonlinear terms that the linearized analytic solution
neglects. For both modes, the magnitude of the differences scales as at least
$\alpha^2$ until they approach numerical roundoff.
\label{fig:AnaLinTest}
}
\end{figure}

The linearized form for the Bondi-Sachs metric for a shell of outgoing
perturbations on a Minkowski background was given in~\cite{bishop05}, though our
choice of notation follows more closely with that used
in~\cite{Reisswig:2006nt}. We can express the solutions in terms of the metric
quantities
\begin{align}
J_{\text{lin}} =& \sqrt{(\ell+2)!/(\ell-2)!}\ ^2Z^{\ell m}\Re
        \left(J_\ell(r)e^{i\nu u}\right), \nonumber\\
U_{\text{lin}} =& \sqrt{\ell(\ell+1)}\ ^1Z^{\ell m}\Re\left(U_\ell(r)e^{i\nu u}\right),
        \nonumber\\
\beta_{\text{lin}} =& \ ^0Z^{\ell m}\Re\left(\beta_\ell(r)e^{i\nu u}\right), \nonumber\\
W_{\text{lin}} =& \ ^0Z^{\ell m}\Re\left(W_\ell(r)e^{i\nu u}\right),
\label{eq:AnaLinAlphabetSoup}
\end{align}
where $\nu$ is a real constant setting the frequency of the perturbations and
$J_\ell(r), U_\ell(r), \beta_\ell(r)$, and $W_\ell(r)$ are all analytic complex
functions of just the radius and $\ell$-mode of the perturbation, given below.
The angular content is expressed through the various $^sZ^{\ell m}$, which are
just linear combinations of the typical SWSHes defined as in~\cite{bishop05}
\begin{align}
^sZ^{\ell m} =& \frac{1}{\sqrt{2}}\left(^sY^{\ell m} + (-1)^m\ ^sY^{\ell-m}
      \right)&\text{for } m>0, \nonumber\\
^sZ^{\ell m} =& \frac{i}{\sqrt{2}}\left((-1)^m\ ^sY^{\ell m} -\ ^sY^{\ell-m}
      \right)&\text{for } m<0, \nonumber\\
^sZ^{\ell0} =& \ ^sY^{\ell0}.
\end{align}
To get the linearized expression for $H_{\text{lin}}$, we can simply take a direct $u$
derivative of $J_{\text{lin}}$. Since these expressions are defined according to the
Bondi metric, with the surface area coordinate $r$ (rather than $\rho$),
$u$ derivatives are taken along curves
of constant $r$. Thus
$H_{\text{lin}}=J_{\text{lin},u}$.

From this, the linearized news function can be expressed as
\begin{align}
\mathcal N_{\text{lin}} =&\Re\left(e^{i\nu u}\lim_{r\to\infty}
    \left(\frac{\ell(\ell+1)}{4}J_\ell - \frac{i\nu r^2}{2}J_{\ell,r}\right)
    + e^{i\nu u}\beta_\ell\right) \nonumber\\
    \times&\sqrt{\frac{(\ell+2)!}{(\ell-2)!}}\ ^2Z^{\ell m}.
\label{eq:AnaLinNews}
\end{align}

Reference~\cite{Reisswig:2006nt} explicitly wrote out the solutions to the linearized
evolution quantities and news function for the $\ell=2$ and $\ell=3$ modes,
which we reproduce here. For $\ell=2$,
\begin{align}
\beta_2 =& B_2, \nonumber\\
J_2(r) =& \frac{24B_2+3i\nu C_{2a}-i\nu^3C_{2b}}{36} + \frac{C_{2a}}{4r}
      - \frac{C_{2b}}{12r^3}, \nonumber\\
U_2(r) =& \frac{-24i\nu B_2+3\nu^2C_{2a}-\nu^4C_{2b}}{36}
      + \frac{2B_2}{r} + \frac{C_{2a}}{2r^2} \nonumber\\
     +&\frac{i\nu C_{2b}}{3r^3}
      + \frac{C_{2b}}{4r^4}, \nonumber\\
W_2(r) =& \frac{24i\nu B_2-3\nu^2C_{2a}+\nu^4C_{2b}}{6} \nonumber\\
     +& \frac{3i\nu C_{2a}-6B_2-i\nu^3 C_{2b}}{3r} \nonumber\\
     -& \frac{\nu^2C_{2b}}{r^2}+\frac{i\nu C_{2b}}{r^3}
      + \frac{C_{2b}}{2r^4}, \nonumber\\
\mathcal N^{2m} =& \Re\left(\frac{i\nu^3C_{2b}}{\sqrt{24}}e^{i\nu u}
    \right)\ ^2Z^{2m},
\label{eq:AnaLinAlphabetSoupL2}
\end{align}
and for $\ell=3$,
\begin{align}
\beta_3 =& B_3, \nonumber\\
J_3(r) =& \frac{60B_3+3i\nu C_{3a}+\nu^4C_{3b}}{180} + \frac{C_{3a}}{10r}
      - \frac{i\nu C_{3b}}{6r^3} - \frac{C_{3b}}{4r^4}, \nonumber\\
U_3(r) =& \frac{-60i\nu B_3+3\nu^2C_{3a}-i\nu^5C_{3b}}{180}
      + \frac{2B_3}{r} + \frac{C_{3a}}{2r^2} \nonumber\\
     -&\frac{2\nu^2C_{3b}}{3r^3} + \frac{5i\nu C_{3b}}{4r^4}
      + \frac{C_{3b}}{r^5}, \nonumber\\
W_3(r) =& \frac{60i\nu B_3-3\nu^2C_{3a}+i\nu^5C_{3b}}{15} \nonumber\\
     +& \frac{i\nu C_{3a}-2B_3+\nu^4C_{3b}}{3r} \nonumber\\
     -& \frac{2i\nu^3C_{3b}}{r^2}-\frac{4i\nu^2C_{3b}}{r^3}
      + \frac{5\nu C_{3b}}{2r^4} + \frac{3C_{3b}}{r^5}, \nonumber\\
\mathcal N^{3m} =& \Re\left(\frac{-\nu^4C_{3b}}{\sqrt{30}}e^{i\nu u}
    \right)\ ^2Z^{3m},
\label{eq:AnaLinAlphabetSoupL3}
\end{align}
where $B_\ell, C_{\ell a}$, and $C_{\ell b}$ are all freely chosen complex constants. Note
that only the values of $C_{\ell b}$ show up in the expression for the news.

For the tests we performed here, we follow a similar setup as
in~\cite{Reisswig:2006nt, Reisswig:2012ka}, where we evolve a system which is a
simple linear combination of the $(2,2)$ and $(3,3)$ modes. Specifically, the
parameter values are $\nu=1$, $B_\ell=0.5i\alpha$, $C_{\ell a}=1.5\alpha$, and
$C_{2b}=-iC_{3b}=0.5\alpha$, where the constant $\alpha$ sets the amplitude of
the resulting news as well as the scale of the linearity of the system. Because
we evolve the entire nonlinear solution, and not just a linearized version, we
expect our results to differ from the analytic solution with differences that
scale as the square of the amplitude, $\alpha^2$.

We place these linearized values of the evolution quantities $(J,W,U,\beta)$ on
a chosen worldtube to serve as the inner boundary values for the volume
evolution. By starting with the worldtube in the Bondi metric, we bypass the
entire inner boundary formalism since we are already starting with the Bondi
metric quantities. To make this test even more demanding, we chose our worldtube
such that its surface area coordinate varies both in time and across the surface,
given by
the formula
\begin{widetext}
\begin{align}
R(u,x,y,z) = 5\left(1+\frac{(-0.42x+0.29y+0.09z)(0.2x+0.1y-0.12z)(0.7x+0.1y-0.3z)
            (0.12x-0.31y-0.5z)}{(x^2+y^2+z^2)^2}\sin{\pi u}\right).
\end{align}
\end{widetext}
We chose this distortion of the surface area coordinate somewhat arbitrarily, ensuring that
it had distortions with modes up through $\ell=4$ as well as a time varying
component with a frequency distinct from that of the linearized perturbation.
This tests the code's ability to distinguish between $H$ and $\Phi$ with the
correct handling of the moving worldtube surface area coordinate, $R$, at least to linear
order. Since this test bypasses the inner boundary formalism, we cannot make
any claim about whether the coordinate radius $\breve r$ of the worldtube is
moving as there is no defined coordinate radius.

The data for $J$ on the initial slice we also read off from
Eq.~(\ref{eq:AnaLinAlphabetSoup}). With the worldtube metric values and initial
slice established, we evolve the full characteristic system. We resolve SWSH modes through
$\ell=8$ with a radinull resolution of 20 grid points and a relative time
integration error tolerance of $10^{-8}$. We test the characteristic evolution
against perturbation amplitudes of $\alpha = (10^{-2}, 10^{-3}, 10^{-4},
10^{-5}, 10^{-6}, 10^{-7}, 10^{-8})$ from $u=0$ to $u=10$. We compute the
difference between the computed news and the analytic results from
Eq.~(\ref{eq:AnaLinNews}), $|\Delta N^{\ell m}| = |N^{\ell m}_{\text{Char}} -
\mathcal N^{\ell m}_{\text{lin}}|$ in Fig.~\ref{fig:AnaLinTest}. Note, we are examining
the news function evaluated at the $\mathscr{I}^+$ coordinates
$(u,\theta,\phi)$, rather than the inertial coordinates $(\tilde u,
\tilde\theta, \tilde\phi)$, because we expect the difference between the two
systems to be a small correction to the linearized values.

From Fig.~\ref{fig:AnaLinTest} we clearly see that when $\alpha\gtrsim10^{-6}$,
$|\Delta N^{\ell m}|$ scales as $\alpha^2$. When $\alpha\lesssim10^{-6}$, the
difference in news rapidly reaches a floor below $10^{-14}$ for the smallest
amplitude perturbations. Modes other than $(2,\pm2)$ and $(3,\pm3)$ all converge
toward $0$ with scaling behavior no worse than $|\Delta N^{\ell m}|
\lessapprox\mathcal{O}(\alpha^2)$ until reaching machine roundoff.
The observed scaling with $\alpha$ matches the expected scaling: we are evolving
the full nonlinear equations but are comparing to an analytic solution of the
linearized equations.

Previous iterations of CCE codes have performed a similar linearized analytic
test~\cite{BabiucEtAl2008, Reisswig:2009rx}. While their choice of
parameters differs slightly from ours, they are most similar to our
$\alpha=10^{-6}$, with inner boundaries at fixed, uniform $R$ worldtube
surfaces.
The error in their news at
the resolutions they tested was worse
than $10^{-10}$, whereas the
error in our news for the $\alpha=10^{-6}$ case is at the order of $10^{-14}$,
hovering just about the error of our numerical roundoff. While comparing our
results to theirs is not exactly a 1-1 comparison, we believe this is evidence
for how effective our code is at resolving the linear case.

\subsection{Teukolsky wave}\label{subsec:TeukWave}

\begin{figure}[t!]
\includegraphics[width=.97\columnwidth]{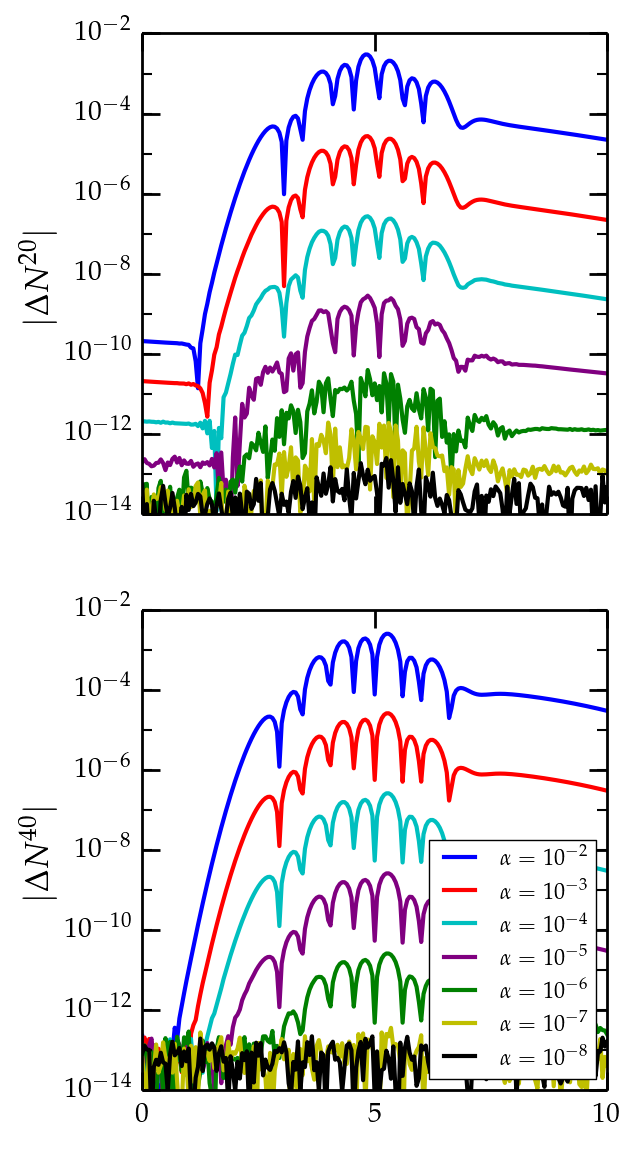}
\caption{
The difference between the numerically evolved news function and the analytic
solution for the Teukolsky wave test of Sec.~\ref{subsec:TeukWave}, for
various amplitudes of the linear perturbation $\alpha$. The $(2,0)$ mode is on the
left and the $(4,0)$ mode on the right. We expect differences of order $\alpha^2$
because we evolve the nonlinear terms that the Teukolsky wave solution neglects.
For both modes, the magnitude of the differences scales as at least $\alpha^2$
until it approaches numerical roundoff.
\label{fig:TeukolskyWave}
}
\end{figure}

A Teukolsky wave is a propagating gravitational wave in the perturbative limit
of Einstein's equations. For outgoing waves the metric has the
form~\cite{teukolsky82}
\begin{align}
d\breve s^2 =& -d\breve t^2 + (1+\breve f_{rr})d\breve r^2
    + 2B\breve f_{r\theta}\breve rd\breve rd\breve\theta \nonumber\\
    +& 2B\breve f_{r\phi}\breve r\sin\breve\theta d\breve rd\breve\phi
    + \left(1+C\breve f_{\theta\theta}^{(1)}
    + A\breve f_{\theta\theta}^{(2)}\right)\breve r^2d\breve\theta^2
    \nonumber\\
    +& 2(A-2C)\breve f_{\theta\phi}\breve r^2\sin\breve\theta
    d\breve\theta d\breve\phi \nonumber\\
    +& \left(1+C\breve f_{\phi\phi}^{(1)} + A\breve f_{\phi\phi}^{(2)}\right)
    \breve r^2\sin^2\breve\theta d\breve\phi^2,
\label{eq:TWaveMetric}
\end{align}
where the functions $\breve f_{ij}$
are known functions of angles listed below,
and the functions $A,B,$ and $C$ are computed from the
freely specifiable function $F(\breve u) = F(\breve t-\breve r)$,
\begin{align}
A =& 3\left(\frac{d_{\breve u}^2F}{\breve r^3}+\frac{3d_{\breve u}F}{\breve r^4}
    + \frac{3F}{\breve r^5}\right), \nonumber\\
B =& -\left(\frac{d_{\breve u}^3F}{\breve r^2}
    + \frac{3d_{\breve u}^2F}{\breve r^3} + \frac{6d_{\breve u}F}{\breve r^4}
    + \frac{6F}{\breve r^5}\right), \nonumber\\
C =& \frac{1}{4}\left(\frac{d_{\breve u}^4F}{\breve r}
    + \frac{2d_{\breve u}^3F}{\breve r^2}+\frac{9d_{\breve u}^2F}{\breve r^3}
    + \frac{21d_{\breve u}F}{\breve r^4}+\frac{21F}{\breve r^5}\right),
\end{align}
where $d_{\breve u}$ is the total derivative with respect to $\breve u$.
The choice of
$F(\breve t-\breve r)$ specifies outward propagating waves, as opposed to
$F(\breve t+\breve r)$ which would generate ingoing waves.

Following~\cite{Fiske2005, Babiuc2005}, we choose the outgoing solution
corresponding to the SWSH $^2Y^{20}$ mode, defining the $\breve f_{ij}$ from
above as
\begin{align}
\breve f_{rr} =& 2-3\sin^2\breve\theta,\ \breve f_{r\theta}
    = -3\sin\breve\theta\cos\breve\theta,\ \breve f_{r\phi}=0, \nonumber\\
    \breve f_{\theta\theta}^{(1)} =& 3\sin^2\breve\theta,\
    \breve f_{\theta\theta}^{(2)} = -1,\ \breve f_{\theta\phi} = 0,
    \nonumber\\
    \breve f_{\phi\phi}^{(1)} =& -\breve f_{\theta\theta}^{(1)},\
    \breve f_{\phi\phi}^{(2)} = 3\sin^2\breve\theta-1,
\end{align}
and defining the profile of the waves with $F(\breve u) = \alpha e^{-\breve
u^2/\tau^2}$, where $\alpha$ and $\tau$ are the amplitude and width of the wave,
respectively. This is slightly different from the choice of $F(\breve u)$ used
in either~\cite{Fiske2005} or~\cite{Babiuc2005}.

Because this solution starts with a metric
that is not in Bondi-Sachs form, this test
utilizes the full inner boundary formalism, in contrast to the linearized
analytic test in Sec.~\ref{subsec:LinAnaSol},
which tests only the characteristic
evolution.
We evaluate the components of
  the metric [see Eq.~(\ref{eq:TWaveMetric})] at a worldtube of
  constant radius, $\breve r_{|\Gamma}$.
  The worldtube treatment in Sec.~\ref{sec:IBF} assumes
  that the metric is given by the 3+1 variables $\breve g_{ij}$, $\breve\alpha$,
  and $\breve\beta^i$ in Cartesian coordinates; we
  obtain these 3+1 Cartesian quantities
  from the spherical components in Eq.~(\ref{eq:TWaveMetric}) in the
  standard way, using $\breve x = \breve r \sin\breve\theta \cos\breve\phi$
  and so on.

Given the metric and its derivatives
evaluated on a worldtube, the inner boundary formalism
creates a correspondence between time and angular coordinates on the worldtube
and at $\mathscr{I}^+$, i.e. $(u=\breve t,\theta=\breve\theta,
\phi=\breve\phi)$. With that in mind, the news function of this waveform at
$\mathscr{I}^+$ is given by the formula~\cite{Babiuc2005}
\begin{equation}
  \mathcal N =
  -\frac{3\sin^2\breve\theta}{4}\partial_{u}^5F(\breve u),
\end{equation}
where $\breve u =
u-\breve r_{|\Gamma}$. For our choice of $F(\breve u)$,
\begin{align}
\mathcal N^{20} =& \alpha\sqrt{\frac{6\pi}{5}}e^{-\breve u^2}\left(120\breve u
    - 160\breve u^3 + 32\breve u^5\right)
\label{eq:TWaveNews}
\end{align}
with all other news modes $\mathcal N^{\ell m\ne20}=0$.
When we compare our computed news
with this analytic news, we do so using the news evaluated on the
coordinates $(u,\theta, \phi)$, rather than the inertial ones $(\tilde
u,\tilde\theta, \tilde\phi)$.

Because this is a solution of the linearized Einstein equations, comparing with
our numerical solution of the full nonlinear equations should yield differences
that scale like $\alpha^2$. Note that even
though we represent the magnitude of the linear perturbation with $\alpha$ in
both this test and the linearized analytic test above, the absolute amplitude
for a given $\alpha$ is not the same for the two tests.
The Teukolsky wave news function here is
over 2 orders of magnitude
larger than the linearized analytic solution for the same value
of $\alpha$.

For our test, the worldtube
  is at a coordinate radius of $\breve
r_{|\Gamma}=5$, and we start
the wave at the origin with a width of $\tau=1$ with
amplitudes $\alpha = (10^{-2}, 10^{-3}, 10^{-4}, 10^{-5}, 10^{-6}, 10^{-7},
10^{-8})$. The CCE code is run to resolve the news up through $\ell=8$ modes with
20 radinull points and a relative time integration error tolerance of
$\approx4\times10^{-6}$. We evolve the system from $u=0$ through $u=10$, which
starts and ends
when the metric is effectively flat.

We show the difference between the numerical evolution and the $(2,0)$ mode of the
analytic news from Eq.~(\ref{eq:TWaveNews}), $|\Delta N^{20}| = |N^{20}_{\text{CCE}} -
\mathcal N^{20}|$ on the left side of Fig.~\ref{fig:TeukolskyWave}. We see for
larger perturbations ($\alpha\gtrsim10^{-6}$) the difference in the news scales
with $\alpha^2$, while for smaller perturbations $(\alpha\lesssim10^{-6})$
$|\Delta N^{20}|$ reaches a floor below $10^{-12}$. For other $\ell=$ even, $m=0$
modes, such as the $(4,0)$ mode plotted on the right half of
Fig.~\ref{fig:TeukolskyWave}, the behavior is similar. Because we chose a
solution with $m=0$, all $m\ne0$ modes of the numerical solution vanish to
numerical roundoff for all $\alpha$.

This behavior is very similar to what we see for the linearized analytic test.
This confirms that our CCE code is consistent with the
linear solution.
Because this test also incorporates the full inner boundary formalism (as
opposed to the linearized analytic test which does not), this also confirms that
to linear order, we reproduce the Bondi metric on the worldtube.

\subsection{Rotating Schwarzschild}

Following the test used in~\cite{Bishop:1997ik}, we generate data
corresponding to the
Schwarzschild metric in Eddington-Finkelstein coordinates with a rotating
coordinate transformation, $\breve\phi\rightarrow\breve\phi+\omega\breve u$, so
the metric is
\begin{align}
d\breve s^2=&-\left(1-\frac{2M}{\breve r}
    - \omega^2\breve r^2\sin^2\breve\theta\right)d\breve u^2
    - 2d\breve ud\breve r\nonumber\\
     +& 2\omega \breve r^2\sin^2\breve\theta d\breve ud\breve\phi
     + \breve r^2\sin^2\breve\theta d\breve\Omega^2,
\end{align}
where $M$ is the mass, $\omega$ is the parameter of the transformation, and
$\breve u$ is the coordinate $\breve u=\breve t-\breve r^*$. For our test, we
chose $M=1$ and $\omega=0.1$. The worldtube has a radius of $\breve r=3M$ and the
solution is evolved from $u=0M$ to $u=0.5M$.
Because the metric is just Schwarzschild in different coordinates, there is no
gravitational radiation. We ran our code with an
$\mathit{absolute}$ time integration error tolerance of $10^{-12}$ and an inertial
coordinate damping parameter of $\kappa=10$. The resulting numerical values of
all the news modes (resolved up through $\ell=8$) are below absolute values of
$10^{-12}$. Because this case uses a spacetime metric that is not in Bondi form
and has a nontrivial angular dependence, it is a full, nonlinear test of our
code (albeit with no time dependence) from the inner boundary formalism through
the extraction of the news function at $\mathscr{I}^+$.

\subsection{Bouncing black hole}
\label{sec:bouncing-black-hole}

\begin{figure}[t!]
\includegraphics[width=.97\columnwidth]{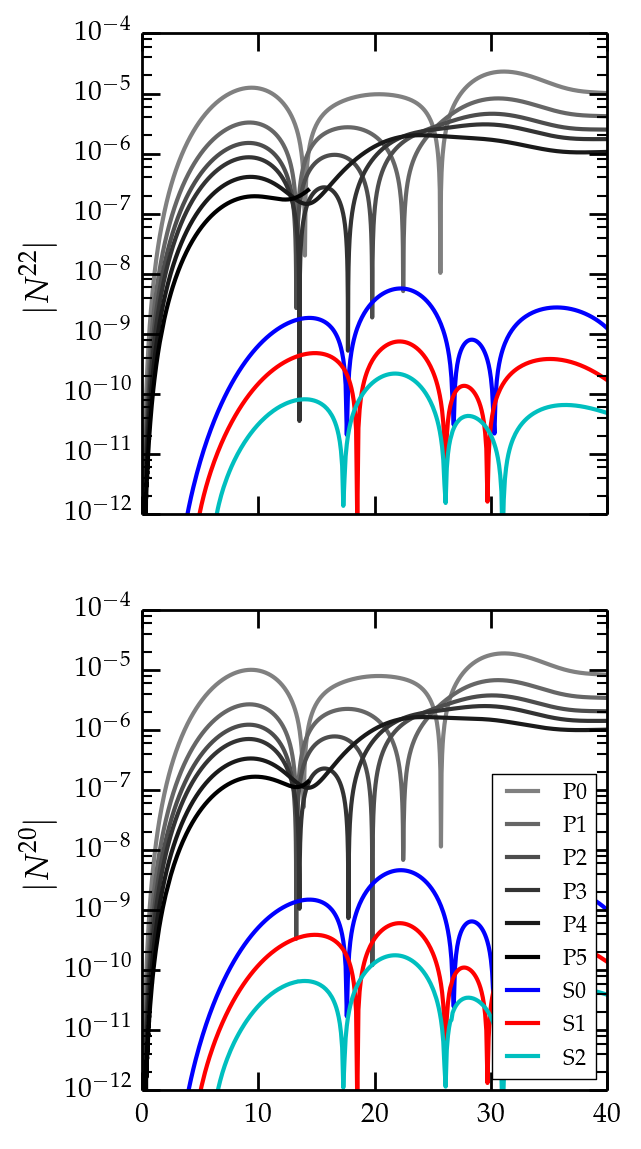}
\caption{
The absolute values of the $(2,2)$ and $(2,0)$ news modes for both the SpEC
(color, resolutions denoted by S0 through S2) and \textsc{Pitt\-Null} (grayscale,
resolutions denoted by P0 through P5) CCE codes for the bouncing black hole test
(Sec.~\ref{sec:bouncing-black-hole}). For this test the news should be zero.
Although both codes are convergent, the SpEC results achieve much smaller errors
than the \textsc{Pitt\-Null} results, especially near the beginning and end of the cycle
as the off-center translation vanishes.
\label{fig:BouncingBH_22_20}
}
\end{figure}

\begin{figure}[t!]
\includegraphics[width=.97\columnwidth]{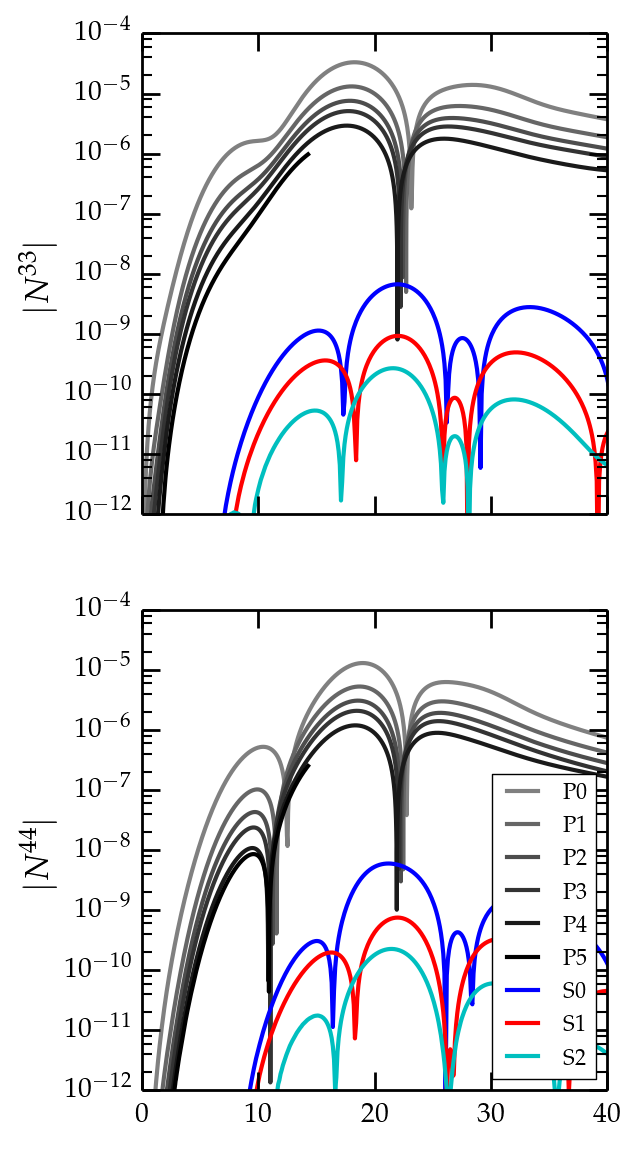}
\caption{
Same as Fig.~\ref{fig:BouncingBH_22_20}, but for the $(3,3)$ and $(4,4)$ modes of
the news. For these modes, the SpEC news is at least two orders of magnitude
smaller than that of \textsc{Pitt\-Null}.
\label{fig:BouncingBH_33_44}
}
\end{figure}

One expected key feature of CCE is its ability to remove gauge effects from the
resulting waveform regardless of the coordinates of the Cauchy metric. We
construct a test similar to those in~\cite{Babiuc2005, Reisswig:2009rx}. We
start with a Schwarzschild black hole and apply a simple time-dependent periodic
coordinate translation on the spacetime. Doing so produces a time-dependent,
periodic metric at the (coordinate-stationary) worldtube, but because this
black hole is not radiating, the news function of this spacetime should be zero;
the goal of this test is to verify that we indeed get zero in this nonlinear,
time-dependent situation.

Specifically, the solution is that of a Schwarzschild black hole with mass $M=1$
in Kerr-Schild coordinates $(\breve t, \breve x, \breve y, \breve z)$, with a
simple oscillating coordinate transformation
\begin{align}
\breve x \rightarrow \breve x + a\sin^4\left(\frac{2\pi \breve t}{b}\right),
\end{align}
where in our test we chose $a=2M$ and $b=40M$. Thus, in the
coordinate frame, which is also the frame of the worldtube, the black hole will
appear to bounce back and forth along the
$\breve x$-axis, but
there is no radiated gravitational wave content. The worldtube is placed at
$\breve r_{\Gamma}=15M$,
which is intentionally very small compared to
  what would be used for a compact binary simulation (typically hundreds of $M$); we chose an artificially small worldtube to produce an extremely difficult
  test of the CCE code. We evolve the system from $u=0M$ to $u=40M$, one full
period of the coordinate oscillation, starting and ending when the coordinates
of the black hole are at the origin.

We performed the characteristic
evolution with our spectral code at three different
resolutions, which we label as S$k$, where $k$ is $(0,1,2)$. We set the resolution at
each level of refinement as follows:
  we retain SWSH modes ${}^sY^{\ell m}$ through
$\ell_{\text{max}}=8+2k$, we use $20+2k$
collocation points in the radinull direction, and the adaptive
  time stepper uses a
relative error tolerance of $3\times10^{-5}\times e^{-k}$
with a maximum step
size of $\Delta u=0.1$. For each resolution, we ran our code on a single core on
the $Wheeler$ cluster at Caltech an Intel Xeon E5-2680,
taking less than $(30, 50, 120)$ minutes for
the (S0, S1, S2) resolutions, respectively.

  For simplicity, we examine the news at $\mathscr{I}^+$ in the coordinates
  $(u,\theta,\phi)$ rather than in the inertial coordinates $(\tilde u,
  \tilde\theta, \tilde\phi)$. Similarly, we
  expand the news into spherical harmonic
modes ${}^{2}Y^{\ell m}(\theta,\phi)$. Since the
news
function is supposed to be zero uniformly, simple coordinate transformations at
$\mathscr{I}^+$ are not expected to affect the overall results presented here.

As a baseline for comparison, we also ran the \textsc{Pitt\-Null} code on the same worldtube data.
We ran \textsc{Pitt\-Null} at multiple resolutions (P0-P5).
  These correspond to a resolution of $(100^3,
200^2, 300^3, 400^3, 600^3, 900^3)$ spatial points and fixed time steps of
$\Delta u=(0.05, 0.025, 0.01667, 0.0125, 0.00833, 0.00556)M$. Because \textsc{Pitt\-Null} takes
significant computational resources at high resolution, we intentionally
terminated the P5 simulation after less than $15M$. During the time that it ran,
that simulation continued trends seen in the lower resolution \textsc{Pitt\-Null}
simulations. The \textsc{Pitt\-Null} resolutions (P0, P1, P2) were run on 24 cores on the
$Wheeler$ cluster at Caltech, taking approximately (850, 2650, 5350) total
CPU hours, respectively, while resolutions (P3, P4, P5) were run on 512 cores on
the BlueWaters cluster, taking approximately (9000, 17000,
24000) total CPU hours, respectively. In the case of P5, that corresponds to the
cost expended on the simulation before we terminated it. This massive
discrepancy on computational costs between the two codes demonstrates the
impressive speed-up achieved by utilizing spectral methods, similar to what was
observed with the previous implementation of this spectral
code~\cite{Handmer:2014, Handmer:2015}.

In Figs.~\ref{fig:BouncingBH_22_20} and ~\ref{fig:BouncingBH_33_44}, we plot
the amplitudes of the $(2,2), (2,0), (3,3)$, and $(4,4)$ modes of the news
  for both codes for
all resolutions for one oscillation period. In
both codes, the amplitude of the $\ell+m=$ odd modes
vanishes except for numerical roundoff,
likely
due to the planar symmetry of the system.
For the $\ell+m=$ even modes the computed
numerical news is nonzero for both codes at finite resolution.

We see in Fig.~\ref{fig:BouncingBH_22_20} that for the $\ell=2$ modes the SpEC
code does a better job than the \textsc{Pitt\-Null} code does at removing the gauge effects
from the news function, at our chosen resolutions. This is especially true
at the beginning and end of the oscillations when the difference between the
shifted coordinates and Schwarzschild is minor.

During the middle of the period, when the coordinate effects on the worldtube
metric are the largest, the difference between the SpEC and \textsc{Pitt\-Null} news in
the $(2,2)$ and $(2,0)$ modes is the smallest. Yet even in this regime, the lowest
resolution SpEC simulation improves on the highest resolution \textsc{Pitt\-Null}
simulation by over an order of magnitude. For the higher order modes, like the
$(3,3)$ or $(4,4)$ modes in Fig.~\ref{fig:BouncingBH_33_44}, the peak errors in
the lowest resolution SpEC results are roughly 2 orders of magnitude better
than those of \textsc{Pitt\-Null}. In all the modes, improving the SpEC CCE resolution
reduces the amplitude of the news, suggesting the remaining errors in the SpEC
results are due to finite numerical resolution, rather than any issue inherent
to the code.

\begin{figure}
\includegraphics[width=.97\columnwidth]{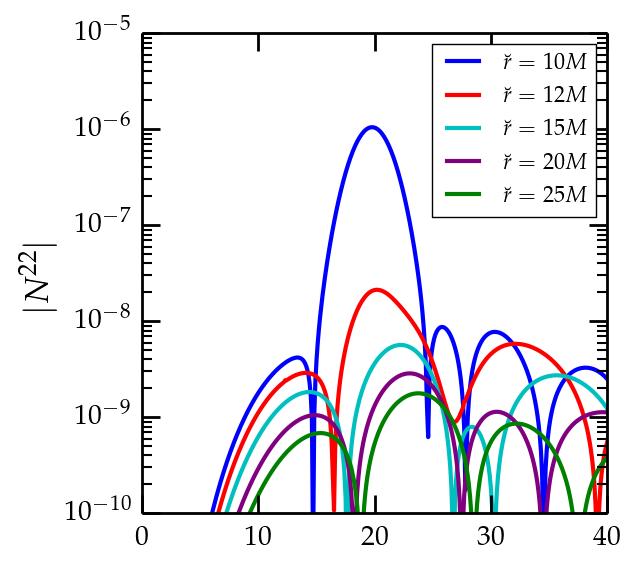}
\caption{
The absolute values of the $(2,2)$ news modes from our SpEC code at the lowest
numerical resolution $S1$, for the bouncing black hole test at different
coordinate worldtube radii $\breve r$. 
\label{fig:BouncingBH_WTR}
}
\end{figure}

This test is a rather extreme test of the code's ability to distinguish
coordinate effects, with the black hole moving an appreciable
fraction of the worldtube's radius in its coordinate frame. We also ran our
code at the lowest resolution on this identical system while placing the worldtube
radius at a series of different coordinate values, $\breve r_{|\Gamma}\in
(10,12,15,20,25)M$, spread quasi-uniformly in $1/\breve r$. In
Fig.~\ref{fig:BouncingBH_WTR}, we plot the amplitude of our code's $(2,2)$ mode for
each of these worldtube radii.

Moving the worldtube to smaller radii raises the error as might be expected;
eventually if the worldtube is close enough to the BH we expect caustics to
form (i.e. radially outward null rays cross paths) and the
characteristic formulation to fail. There is a clear
convergence of this error to zero as we move the worldtube farther away and the
relative size of the coordinate transformation of the bouncing BH shrinks.
 
\subsection{Gauge wave}\label{subsec:GaugeWave}

\begin{figure}[t!]
\includegraphics[width=.97\columnwidth]{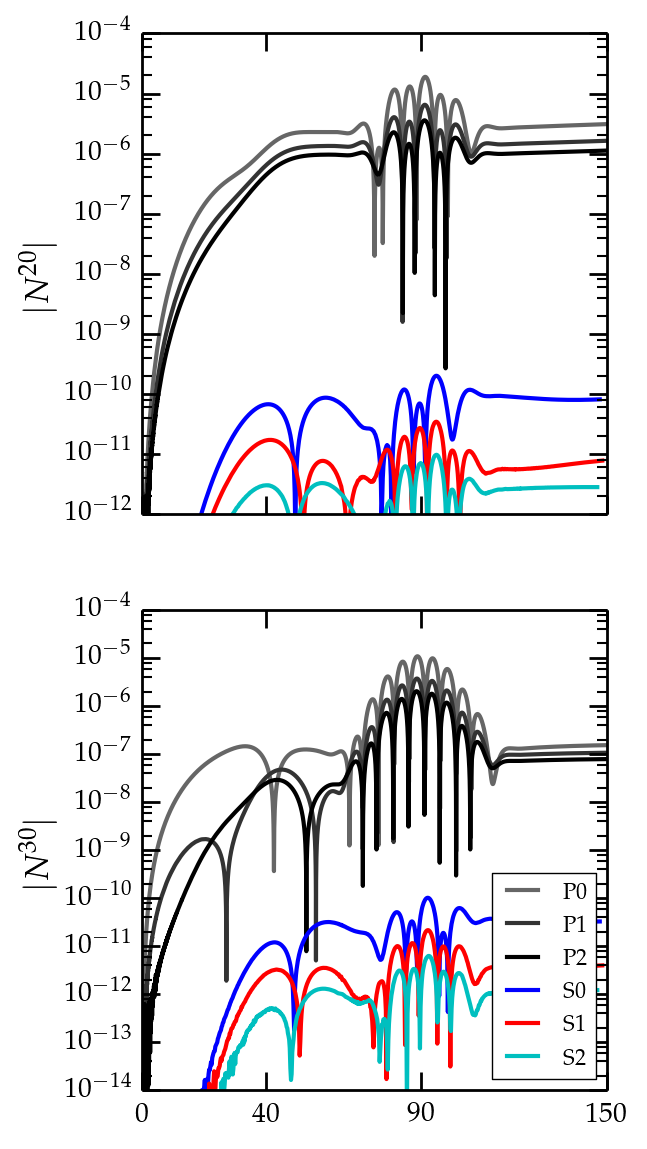}
\caption{
The amplitude of the $(2,0)$ and $(3,0)$ modes of the news for SpEC (color) and
\textsc{Pitt\-Null} (grayscale) CCE codes, for the gauge wave test
(Sec.~\ref{subsec:GaugeWave}). The center of the coordinate shift off-center
occurs around $u=40M$ while the peak of the gauge wave propagates to
$\mathscr{I}^+$ at $u=90M$. For this test, the news should be zero. At all
times, the SpEC code is orders of magnitude more accurate than the \textsc{Pitt\-Null}
code.
\label{fig:GaugeWave}
}
\end{figure}

The bouncing black hole test is a measure of the code's ability to remove
coordinate effects resulting from simple translations; we now introduce a test
to examine the code's ability to distinguish between outgoing gravitational
waves and gauge waves propagating along null slices. To generate this gauge
wave, we construct a metric similar to
that introduced by Eq.~(5.2)
in Ref~\cite{Zhang:2012ky}, except modified for an outward propagating gauge
transformation.  Starting with the Schwarzschild metric in ingoing
Eddington-Finkelstein coordinates, we apply the
transformation of $\breve v=\breve t+\breve r+F(\breve t-\breve r)/\breve r$
where $F(\breve u)$ is an arbitrary function. The line element is
\begin{widetext}
\begin{align}
d\breve s^2 =& -\left(1-\frac{2M}{\breve r}\right)\left(1
    + \frac{d_{\breve u}F}{\breve r}\right)^2d\breve t^2
    + 2\left(1+\frac{d_{\breve u}F}{\breve r}\right)
    \left(\frac{2M}{\breve r} + \left(1-\frac{2M}{\breve r}\right)
    \left(\frac{d_{\breve u}F}{\breve r}
    + \frac{F}{\breve r^2}\right)\right)d\breve td\breve r \nonumber\\
    +& \left(1-\frac{d_{\breve u}F}{\breve r}-\frac{F}{\breve r^2}\right)
    \left(1+\frac{2M}{\breve r}+\left(1-\frac{2M}{\breve r}\right)
    \left(\frac{d_{\breve u}F}{\breve r}
    + \frac{F}{\breve r^2}\right)\right)d\breve r^2 + \breve r^2d\breve\Omega^2.
\end{align}
\end{widetext}
Here $M$ is the mass of the black hole and $d_{\breve u}$ is the total
derivative with respect to $\breve u$. For the test, we set
$M=1$ and we chose $F$
to be a sine-Gaussian,
\begin{align}
F(\breve u)=\alpha\sin\left(w\breve u+p_0\right)e^{-\frac{(\breve u
    - \breve u_0)^2}{k^2}}.
\end{align}
Here $\alpha$ is the amplitude of the gauge wave, $w$ is the frequency, $p_0$
is the initial phase offset, $\breve u_0$ is the time when the peak is at the
origin, and $k$ is its characteristic width. For our test, we choose $\alpha=M$,
$w=0.5/M$, $p_0=0.01$, $\breve u_0=40M$, and $k=10$.

Because this system is spherically symmetric, most of the terms in the
evolution equations are trivially zero.
In order to make the test more stringent and to generate nonzero terms in the
  evolution equations, we also apply an additional translation to displace the center of the
black hole from the center of the worldtube. The translation used is
\begin{align}
\breve z\rightarrow\breve z+2\left(1-e^{-(\breve t/40)^4}\right).
\end{align}
By moving the system entirely along the $\breve z$-axis, we expect only $m=0$
modes to be excited. We choose the worldtube radius to be $\breve
r_{\Gamma}=50M$. Our gauge wave is configured so that the peak will propagate
outwards and pass through this worldtube at $\breve t=90M$.

We ran our SpEC CCE code at three different resolutions, S$k$, for $k=(0,1,2)$.
This corresponds to angular resolution of $\ell_{\text{max}}=8+2k$, radinull
resolution of $20+2k$ and $\mathit{absolute}$ time integration error tolerance
of $10^{-12}e^{-k}$. The three resolutions, (S0, S1, S2) were run on a single
core on Caltech's Wheeler cluster for approximately $(35, 75, 165)$
minutes.

\textsc{Pitt\-Null} CCE was also run at three resolutions, $P0$-$P2$ corresponding to a
finite differencing grid with $(100^3, 200^3, 300^3)$ spatial points and fixed
times steps of size $\Delta u=(0.05, 0.025, 0.01667)M$. Each resolution was run on
256 cores on the BlueWaters cluster, costing approximately
(1100, 3200, 6000) CPU hours.

In Fig.~\ref{fig:GaugeWave}, we plot the amplitude of the $(2,0)$ and $(3,0)$
news modes for both codes and in both modes. We expect the news to be zero
because the solution is merely Schwarzschild in moving coordinates.
At all times both codes show convergence toward zero, with SpEC several orders
of magnitude below \textsc{Pitt\-Null}.
In the SpEC results, at the times corresponding to the coordinate shift,
we see the amplitude of the news
is noticeably smaller than seen in the
bouncing black hole test, consistent with the larger worldtube radius used in
this test. The passing gauge wave also leaves an imprint on the news that is
appropriately vanishing with resolution.

Examining the higher $\ell$ modes yields a similar picture for both codes just
at slightly decreasing amplitudes, as seen in the
  right panel of Fig.~\ref{fig:GaugeWave}.
Also, as expected by the axisymmetry of the
setup for this test, both codes produce zero news to numerical
  roundoff for all $m\ne0$ modes.

\section{Conclusion}\label{sec:conc}

In this paper, we have detailed the implementation of our spectral CCE code
as a means of extracting gravitational wave information from
an interior Cauchy evolution of a relativistic
system. We summarized the full theoretical framework CCE along
with discussion of the changes made to the previous version of the
code~\cite{Handmer:2014, Handmer:2015}. In particular, beyond bug fixes and
miscellaneous alterations to the code, we have improved the numerical treatment
of the poles contained within the $Q$, $W$, and $H$ evolution equations,
switched the time stepper from fixed step size to a fifth order adaptive, changed
the representation of the inertial coordinates at $\mathscr{I}^+$ for better
spectral handling. All of these cumulative effects lead to a more robust and
accurate code than before. This paper also clarifies a number of analytic
subtleties and paper typos present within~\cite{Handmer:2014, Handmer:2015}.

We applied our code to a number of analytic test cases in order to examine its
efficacy to extract the correct gravitational wave content from the worldtube
data. In the pair of linearized test cases, the code successfully reproduces the
analytic solution to linear order, with their differences scaling as expected
(i.e. scaling by the nonlinear terms unaccounted for by the linear
approximations). In these two tests, the code is ultimately limited by the
numerical truncation limit of using double precision. A third test, a
Schwarzschild black hole in a rotating coordinate frame, is a full nonlinear
test of the code with a straightforward vanishing solution. Similar to the
linear tests, the code resolves this solution up to numerical truncation limits.

The other two tests, the bouncing black hole and the gauge wave, are more
rigorous tests of the code's capability of eliminating gauge effects from the
final output, and are successful at doing so. For these tests, the errors are
small and convergent with resolution. Furthermore, as the worldtube boundary is
placed farther from the black hole, less resolution is needed to attain a given
level of error.

Overall, this version of the code shows marked
improvements from the previous standards set by the \textsc{Pitt\-Null} code.
In both the
bouncing black hole and gauge wave tests, we ran \textsc{Pitt\-Null} at a series of
different resolutions to serve as an independent comparison. The resulting news
output from our code, for tests where the news should be zero, was orders of
magnitude smaller than that of \textsc{Pitt\-Null}. In addition, we still observe the
computational speed-up of our code by a factor of $>100$ that had been noted
in~\cite{Handmer:2014, Handmer:2015}.

Our current goal is to run our CCE code on the catalog of SpEC
waveforms~\cite{SXSCatalog2018,SXSCatalog}. In future work, we plan to couple the CCE code
to run concurrently with the SpEC Cauchy evolution. Then CCE would not
have to be run as a separate postprocessing step to generate
the final waveforms. We would then like to follow that with
Cauchy-characteristic matching (CCM)~\cite{Bishop1998}, whereby information
from the Bondi metric is fed back into the Cauchy domain as both
the Cauchy and the characteristic systems are jointly evolved. The
characteristic evolution would then couple directly with the Cauchy
evolution, removing the need for boundary conditions
at the artificial outer boundary of the Cauchy domain.
While a previous code has successfully performed CCM
in the linearized case, they were unable to stably run it for the
general case~\cite{Szilagyi00a}.

%%%%%%%%%%%%%%%%%%%%%%%%%%%%%%%%%%%%%%%%%%%%%%%%%%%%%%%%%%%%%%%%%%%%%%%%%%%%%%%
% Acknowledgments
%%%%%%%%%%%%%%%%%%%%%%%%%%%%%%%%%%%%%%%%%%%%%%%%%%%%%%%%%%%%%%%%%%%%%%%%%%%%%%%
\section{Acknowledgments}
We thank
Leo Stein, Casey Handmer, Harald Pfeiffer, Jeff Winicour, and
Saul Teukolsky for helpful advice and many useful discussions about
various aspects of this project.
This work was supported in part by the Sherman Fairchild Foundation
and by NSF Grants No. PHY-1708212 and No. PHY-1708213 at Caltech.
Computations were performed on the Wheeler cluster at Caltech, which
is supported by the Sherman Fairchild Foundation and by Caltech, and
on NSF/NCSA BlueWaters under allocation NSF PRAC-1713694. This
research is part of the BlueWaters sustained-petascale computing
project, which is supported by the National Science Foundation (Awards No.
OCI-0725070 and No. ACI-1238993) and the state of Illinois. BlueWaters is
a joint effort of the University of Illinois at Urbana-Champaign and
its National Center for Supercomputing Applications.

\appendix

\section{Spin-Weighted Spherical Harmonics}
\label{subsec:SWSH}

Spin-weighted spherical harmonics (SWSH) are a generalization of the typical
spherical harmonics by introducing spin-weight raising ($\eth$) and lowering
operators ($\bar\eth$)~\cite{NewmanPenrose1966, Goldberg1967}. These derivative
operators are defined by contracting the dyads with the angular derivative
operator.
For any spin-weighted scalar quantity
  $v=q_1^{A_1} \ldots q_2^{A_2} v_{A_1 \ldots A_n},$ where each $q_i$ may be
  either $q$ or $\bar{q}$, we define the spin-weighted derivatives,
  \begin{align}
    \eth v = q_1^{A_1} \ldots q_n^{A_n} q^B D_B v_{A_1 \ldots A_n},\label{eq:Eth}\\
    \bar{\eth} v = q_1^{A_1} \ldots q_n^{A_n} \bar{q}^B D_B v_{A_1 \ldots A_n},
    \label{eq:EthBar}
  \end{align}
where $D$ is the angular covariant derivative on the unit sphere.
By contracting these dyads with the tensor component gives the spin-weighted
version of the quantities, computed above in Eqs.~(\ref{eq:Jdef})-(\ref{eq:Qdef}). The
dyads contracted with a given quantity determine its spin weight, with $+1$ for
each $q^A$, $-1$ for each $\bar q^A$. For example, the spin weight of $\eth\bar
J=\frac{1}{2}\partial_Ah_{BC}q^A\bar q^B\bar q^C$ is $-1$. Thus we see that $(K,
\beta, W)$ have spin weight of 0, $(Q, U)$ have spin weight 1, and $(J, H,
\Phi)$ have spin weight 2.

Now we can also express $\eth$ as a complex spherical derivative operator on a
given quantity $F$ with a spin weight of $s$, and for our choice of dyad given
  in Eq.~(\ref{eq:Dyad}),
\begin{align}
\eth F =& -\sin^s\theta\left(\frac{\partial}{\partial\theta}
    + \frac{i}{\sin\theta}\frac{\partial}{\partial\phi}\right)(\sin^{-s}\theta F),
    \\
\bar\eth F =&-\sin^{-s}\theta\left(\frac{\partial}{\partial\theta}
    - \frac{i}{\sin\theta}\frac{\partial}{\partial\phi}\right)(\sin^s\theta F).
\end{align}
While \textsc{Pitt\-Null} used a finite difference formulation for computing these
derivatives~\cite{Gomez:1997}, our code will make use of how $\eth$ acts on
individual SWSH modes,
\begin{align}
\eth~^sY^{\ell m} =& \sqrt{(\ell-s)(\ell+s+1)}~^{s+1}Y^{\ell m}
    \label{eq:eth} \\
\bar\eth~^sY^{\ell m} =& -\sqrt{(\ell+s)(\ell-s+1)}~^{s-1}Y^{\ell m}
    \label{eq:beth}
\end{align}
With this, we can start from the regular spherical harmonics ($s=0$) and build
up the SWSH modes for arbitrary spin weight.

And just like regular spherical harmonics, we can take an arbitrary
spin-weighted function of and decompose into spectral coefficients with the use of
the expression of orthonormality of the SWSHes over the unit sphere,
\begin{align}
\int_{S^2}~^sY^{\ell m}~\overline{^sY^{\ell'm'}}d\Omega =&
    \delta_{\ell\ell'}\delta_{mm'},
\label{eq:SWSHOrthonormality}
\end{align}
where $d\Omega$ is the area element of the unit sphere $S^2$.
Thus, given a spin-weighted quantity, we can decompose it as a sum of SWSH modes
and take $\eth$ and $\bar\eth$ derivatives by applying the properties of
Eqs.~(\ref{eq:eth}) and (\ref{eq:beth}) to the spectral coefficients.

Last, we list some basic, useful properties of SWSHes:
\begin{enumerate}[label=(\roman*)]
\item It is only possible to add together spin-weighted quantities of identical spin weight.
\item The spin weight of a product of two SWSHes is the sum of their individual
spin weights.
\item Because typical spherical harmonics are more generally SWSHes of
spin weight 0, SWSHes inherit the same mode properties of spherical harmonics
(i.e. $\ell\geq0,|m|\leq\ell$).
\item In addition, the spin weight serves as a lower bound on possible $\ell$
modes, $\ell\geq|s|$.
\item The $\eth$ and $\bar\eth$ operators do not commute as, given spin-weighted
quantity $F$ of spin $s$, $\bar\eth\eth F = \eth\bar\eth F + 2sF$.
\end{enumerate}

We utilize two external code packages to assist with the numerical
implementation for the angular basis function,
{\tt Spherepack}~\cite{Boyd1989,spherepack-home-page} for the standard spherical
harmonics and {\tt Spinsfast}~\cite{Huffenberger:2010} for the SWSHes. In
particular, we use {\tt Spherepack} primarily during the inner boundary formalism
and partially during $\mathscr{I}^+$ extraction, while we use {\tt Spinsfast} during the
volume evolution and $\mathscr{I}^+$ extraction.

\section{Nonlinear Evolution Equations}\label{subsec:NonLinear}

The full system of nonlinear equations appears below. The equations are the
radinull equations on the null hypersurface for a given time slice.
Reference~\cite{Bishop1996} computed these full nonlinear expressions and first
expressed them as SWSH quantities in~\cite{Bishop:1997ik}, although we
follow~\cite{Handmer:2014} by writing them in terms of the compactified
coordinate $\rho$,
\begin{widetext}
\begin{align}
\beta_{,\rho} =& \frac{\rho(1-\rho)}{8}\left(J_{,\rho} \bar{J}_{,\rho}
    - K_{,\rho}^2\right),
\label{eq:betaEvo}
\\
\left(r^2Q\right)_{,\rho} =& \frac{1}{(1-\rho)^2}\bigg[R^2\rho^2\left(\vphantom{\frac{1}{2K^2}}
        2\eth\beta_{,\rho}-K\eth K_{,\rho}-K\bar{\eth}J_{,\rho}
    + \eth\left(\bar JJ_{,\rho}\right) + \bar\eth\left(JK_{,\rho}\right)
    - J_{,\rho}\bar\eth K \right. \nonumber\\
   &\left. \hphantom{\frac{1}{(1-\rho)^2}}\vphantom{\bar J^2}
        + \frac{1}{2K^2}\left(\eth\bar J(J_{,\rho}-J^2\bar J_{,\rho})
        + \eth J(\bar J_{,\rho}-\bar J^2J_{,\rho}) \right)\right)\bigg] \nonumber\\
   +& \frac{1}{(1-\rho)^3}\left(-4R^2\rho\eth\beta\right),
\label{eq:QEvo}
\\
U_{,\rho} =& \frac{e^{2\beta}}{R\rho^2}\left(KQ-J\bar Q\right),
\label{eq:UEvo}
\\
\mathcal{R} =& 2K - \eth\bar\eth K
    + \frac{1}{2}\left(\bar\eth^2J+\eth^2\bar J\right)
    + \frac{1}{4K}\left(\bar\eth\bar J\eth J - \bar\eth J\eth\bar J\right),
\\
\left(r^2W\right)_{,\rho} =& \frac{1}{(1-\rho)^2} \left(\vphantom{\frac{R^2}{2}}
    - R + \frac{R^2\rho^2}{4}(\eth\bar U_{,\rho}+\bar\eth U_{,\rho})
        -e^{-2\beta}\frac{R^3\rho^4}{8}(2KU_{,\rho}\bar U_{,\rho} + J\bar U_{,\rho}^2
        +\bar JU_{,\rho}^2) \right.\nonumber\\
    &\hphantom{\frac{1}{(1-\rho)^2}}\left.
    + \frac{Re^{2\beta}}{2}\left(\mathcal{R} - 2K(\eth\beta\bar\eth\beta+\eth\bar\eth\beta)
        +J\bar\eth\beta^2+\bar J\eth\beta^2
        -\eth\beta(\bar\eth K-\eth\bar J)
        -\bar\eth\beta(\eth K-\bar\eth J) \right.\right.\nonumber\\
        &\left.\left.\hphantom{\frac{1}{(1-\rho)^2}+ Re^{2\beta}}
        +J\bar\eth^2\beta+\bar J\eth^2\beta\right)
        \vphantom{\frac{Re^{2\beta}}{2}}\right)\nonumber\\
    +& \frac{1}{(1-\rho)^3}\left(R^2\rho(\eth\bar U + \bar\eth U)\right).
\label{eq:WEvo}
\end{align}

The evolution equation of $J$ is given by $H=J_{,u|r=\text{const}}$,
\begin{align}
(rH)_{,\rho} - \frac{rJ}{2}(H\bar T+\bar HT) = H_A
    + \frac{H_{B1} + H_{B2} + H_{B3} + H_{B4}}{1-\rho}
    + \frac{H_C}{(1-\rho)^2},
\label{eq:HEvo}
\end{align}
where
\begin{align}
T =& \left(J_{,\rho} - \frac{JK_{,\rho}}{K}\right), \label{eq:TOrGamma}\\
H_A =& (1-\rho)J_{,\rho} + \frac{R}{2}\rho^2W_{,\rho}J_{,\rho}
    + \frac{\rho}{2}(1-\rho+R\rho W)J_{,\rho\rho} - 4J\beta_{,\rho}, \\
H_{B1} =& \frac{R\rho}{4}\left((6-4\rho)WJ_{,\rho} - 16JW\beta_{,\rho}
    - \eth J\bar U_{,\rho} - \bar\eth JU_{,\rho} - 2K\eth U_{,\rho}
    - J_{,\rho}\left(\eth\bar U + \bar\eth U\right)
    + J\left(\bar\eth U_{,\rho} - \eth\bar U_{,\rho}\right)\right), \\
H_{B2} =& \frac{R\rho}{4}\left(\left(\bar U\eth J+U\bar\eth J\right)\left(J\bar J_{,\rho}
    - \bar JJ_{,\rho}\right) - 2\bar U\eth J_{,\rho} - 2U\bar\eth J_{,\rho}
    \right. \nonumber\\
    &\hphantom{\frac{R\rho}{4}}+ \left.2\left(KJ_{,\rho}-JK_{,\rho}\right)\left(\bar U\eth K
    + U\bar\eth K + K(\bar\eth U - \eth\bar U) + J\bar\eth\bar U - \bar J\eth U\right)
    \right), \\
H_{B3} =& \frac{e^{2\beta}}{2\rho}\left((2+J\bar J)\left(\eth^2\beta+\eth\beta^2\right)
    + J^2\left(\bar\eth^2\beta+\bar\eth\beta^2\right)
    - 2JK\left(\eth\bar\eth\beta+\bar\eth\beta\eth\beta\right)
    + J\left(\eth K\bar\eth\beta - \eth\beta\bar\eth K + \eth\bar J\eth\beta\right)
    \right.\nonumber\\
    &\left.\hphantom{\frac{e^{2\beta}}{2\rho}}+ \bar J\eth J\eth\beta
    + K\left(\bar\eth J\eth\beta - \eth J\bar\eth\beta - 2\eth K\eth\beta
    \right)\right), \\
H_{B4} =& \frac{e^{-2\beta}R^2\rho^3}{8}\left((2+J\bar J)U_{,\rho}^2
    + 2JKU_{,\rho}\bar U_{,\rho} + J^2\bar U_{,\rho}^2\right), \\
H_C =& -\frac{R}{2}\left(2K\eth U + \eth J\bar U + \bar\eth JU - J\bar\eth U
    + J\eth\bar U\right).
\end{align}
\end{widetext}

\section{Paper Definition Key}\label{subsec:Key}

Here we define the quantities we use in the paper for ease of reference.

\begin{align}
\breve\alpha&: \text{Lapse function in Cauchy metric} \nonumber\\
\beta^{\breve\i}&: \text{Shift vector in Cauchy metric} \nonumber\\
\beta&: \text{Time-time part of metric in Bondi form, Eq.~(\ref{eq:BondiMetric})}\nonumber\\
\eth, \bar\eth&: \text{Angular derivative operators, Eqs.~(\ref{eq:Eth})-(\ref{eq:EthBar})} \nonumber\\
\Gamma&: \text{Worldtube hypersurface}\nonumber\\
g_{\mu\nu}&: \text{Metric in Bondi form, Eq.~(\ref{eq:BondiMetric})}\nonumber\\
\breve g_{\breve\mu\breve\nu}&: \text{Cauchy metric} \nonumber\\
\hat g_{\hat\mu\hat\nu}&=\ell^2g_{\mu\nu}: \text{Compactified metric in Bondi form,
    Eq.~(\ref{eq:BondiMetricCompact})}\nonumber\\
\tilde g_{\tilde\mu\tilde\nu}&=\omega^2\hat g_{\hat\mu\hat\nu}: \text{Conformal metric in Bondi form,
    Eq.~(\ref{eq:BondiMetricConformal})}\nonumber\\
H&: \text{Time derivative of $J$ in Bondi frame, Eq.~(\ref{eq:Hdefinition})}\nonumber\\
h_{AB}&: \text{Angular part of metric in Bondi form, Eq.~(\ref{eq:BondiMetric})} \nonumber\\
\mathscr{I}^+&: \text{Future null infinity} \nonumber\\
J&=\frac{1}{2}h_{AB}q^Aq^B: \text{Spin-weighted angular metric function} \nonumber\\
K&=\sqrt{1+J\bar J}: \text{Auxiliary angular variable} \nonumber\\
\ell&=1/r: \text{Compactified surface-area coordinate}\nonumber\\
\ell^{\breve\mu}&: \text{Worldtube null generator, Eq.~(\ref{eq:WTNullGenerator})} \nonumber\\
\bar\lambda&: \text{Worldtube affine radinull parameter} \nonumber\\
N&: \text{News function, Eq.~(\ref{eq:NewsFunction})} \nonumber\\
n^{\breve\mu}&: \text{Timelike unit vector at worldtube, Eq.~(\ref{eq:WTTimeVector})} \nonumber\\
\hat n^\mu&: \text{Compactified Bondi generator at $\mathscr{I}^+$,
    Eq.~(\ref{eq:CompactifiedGenerator})} \nonumber\\
\tilde n^\mu&: \text{Conformal Bondi generator at $\mathscr{I}^+$,
    Eq.~(\ref{eq:ConformalGenerator})} \nonumber\\
\Phi&: \text{Time derivative of $J$ in affine frame, Eq.~(\ref{eq:EvolutionEquationJ})} \nonumber\\
Q_A&: \text{Radial derivative of $U^A$, Eq.~(\ref{eq:QaDefinition})} \nonumber\\
Q&=Q_Aq^A: \text{Spin-weighted radial derivative of $U^A$} \nonumber\\
q_{A}&: \text{Complex dyad, Eq.~(\ref{eq:ComplexDyads})} \nonumber\\
q_{AB}&: \text{Unit sphere metric} \nonumber\\
R&=r_{|\Gamma}: \text{Radius of worldtube,
    Eq.~(\ref{eq:WorldTubeBondiRadius})}\nonumber\\
\mathcal{R}&: \text{Curvature scalar for angular metric, Eq.~(\ref{eq:Rcurve})} \nonumber\\
r&: \text{Surface-area coordinate} \nonumber\\
\rho&=\frac{r}{R+r}: \text{Compactified surface-area coordinate} \nonumber\\
\breve r&: \text{Radius of worldtube in Cauchy coordinates} \nonumber\\
s^{\breve\mu}&: \text{Spatial outgoing
  unit normal to $\Gamma$,
    Eq.~(\ref{eq:WTOutwardNormal})} \nonumber\\
\breve t&: \text{Time coordinate in Cauchy metric}
\nonumber\\
u&: \text{Retarded time coordinate} \nonumber\\
\tilde u&: \text{Conformal Bondi time coordinate, Eq.~(\ref{eq:ConformalGenerator})}
\nonumber\\
U^A&: \text{Angular shift part of metric in Bondi form, Eq.~(\ref{eq:BondiMetric})} \nonumber\\
U&=U^Aq_A: \text{Spin-weighted angular shift} \nonumber\\
W&: \text{Mass aspect of metric in Bondi form, Eq.~(\ref{eq:BondiMetric})} \nonumber\\
\Omega&: \text{A conformal factor at $\mathscr{I}^+$,
  Eq.~(\ref{eq:OmegaDefinition})}\nonumber\\
d\Omega&: \text{Unit sphere area element} \nonumber\\
\omega&: \text{A conformal factor at $\mathscr{I}^+$,
    Eq.~(\ref{eq:ConformalOmega})}\nonumber\\
x^\alpha&=(u,r,\theta,\phi): \text{Coordinates of $g_{\mu\nu}$, Eq.~(\ref{eq:BondiMetric})} \nonumber\\
\breve x^{\breve\alpha}&:
\text{Coordinates of Cauchy metric
  $\breve g_{\breve\mu\breve\nu}$} \nonumber\\
\bar x^{\bar\alpha}&:
    \text{Coordinates of $\bar g_{\bar\mu\bar\nu}$, Eq.~(\ref{eq:NullMetricDef})} \nonumber\\
    \hat x^{\hat \alpha}&=(u,\ell,\theta,\phi): \text{Coordinates of $\hat g_{\hat\mu\hat\nu}$, Eq.~(\ref{eq:BondiMetricCompact})} \nonumber\\
\tilde x^{\tilde\alpha}&: \text{Coordinates of $\tilde g_{\tilde\mu\tilde\nu}$, Eq.~(\ref{eq:BondiMetricConformal})} \nonumber
\end{align}

\bibliography{References/References}

\end{document}